\documentclass{aa}

\usepackage[varg]{txfonts}
\usepackage{natbib}
\usepackage{graphicx}
\usepackage[switch]{lineno}

\defcitealias{Joubaud19}{paper I}
\newcommand{\g}{\ensuremath{\gamma}\xspace}
\newcommand{\halpha}{\ensuremath{\rm{H}_{\alpha}}\xspace}
\newcommand{\wco}{\ensuremath{W_{\rm{CO}}}\xspace}
\newcommand{\hd}{H$_2$\xspace}
\newcommand{\hi}{\ion{H}{i}\xspace}
\newcommand{\hii}{\ion{H}{ii}\xspace}
\newcommand{\nh}{\ensuremath{N_{\rm{H}}}\xspace}
\newcommand{\nhi}{\ensuremath{N_{\ion{H}{i}}}\xspace}
\newcommand{\nhii}{\ensuremath{N_{\ion{H}{ii}}}\xspace}
\newcommand{\nhd}{\ensuremath{N_{\rm{H}_2}}\xspace}
\newcommand{\nhdnm}{\ensuremath{N_{\rm{H}_{\rm{DNM}}}}\xspace}

\newcommand{\qlis}{\ensuremath{q_{\rm{LIS}}}\xspace}

\newcommand{\qco}{$q_{\rm{CO}}$\xspace}
\newcommand{\qhii}{$q_{\ion{H}{ii}}$\xspace}
\newcommand{\qdnm}{\ensuremath{q_{\rm{DNM}}}\xspace}
\newcommand{\qsb}{\ensuremath{q_{\rm{SB}}}\xspace}
\newcommand{\yhi}{\ensuremath{y_{\ion{H}{i}}}\xspace}
\newcommand{\yco}{\ensuremath{y_{\rm{CO}}}\xspace}
\newcommand{\yhii}{\ensuremath{y_{\ion{H}{ii}}}\xspace}
\newcommand{\ydnm}{\ensuremath{y_{\rm{DNM}}}\xspace}
\newcommand{\yiso}{\ensuremath{y_{\rm{iso}}}\xspace}
\newcommand{\qcounit}{$10^{20}$\,cm$^{-2}$\,K$^{-1}$\,km$^{-1}$\,s\xspace}  
\newcommand{\qhiiunit}{$10^{20}$\,cm$^{-2}$\xspace}
\newcommand{\qdnmTunit}{$10^{25}$\,cm$^{-2}$\xspace} 
\newcommand{\yhTunit}{$10^{-26}$\,cm$^2$\xspace} 
\newcommand{\ycoTunit}{$10^{-6}$\,K$^{-1}$\,km$^{-1}$\,s\xspace} 
\newcommand{\yisoTunit}{$10^{-6}$\xspace} 

\newcommand{\tspin}{\ensuremath{T_{\rm{S}}}\xspace}

\newcommand{\percc}{cm$^{-3}$\xspace}
\newcommand{\persqcm}{cm$^{-2}$\xspace}
\newcommand{\kmpers}{km\,s$^{-1}$\xspace}
\newcommand{\southloop}{South Loop\xspace}
\newcommand{\northrim}{North Rim\xspace}
\newcommand{\eastshell}{East Shell\xspace}
\newcommand{\westrim}{West Rim\xspace}
\newcommand{\bsky}{$B_{\rm sky}$\xspace}

\newcommand{\muG}{\ensuremath{\mu\rm{G}}\xspace}
\def\Fermi{\textit{Fermi}\xspace}
\def\Planck{\textit{Planck}\xspace}

\def\IRAS{\textit{IRAS}\xspace}

\begin{document}

\title{The cosmic-ray content of the Orion-Eridanus superbubble}

\author{T. Joubaud\inst{1}
\and I. A. Grenier\inst{1}
\and J. M. Casandjian\inst{1}
\and T. Tolksdorf\inst{2}
\and R. Schlickeiser\inst{2}}

\institute{
\inst{1}~Laboratoire AIM, CEA-IRFU/CNRS/Universit\'e de Paris, D\'epartement d’Astrophysique, CEA Saclay, F-91191 Gif sur Yvette, France \\
\inst{2}~Institut für Theoretische Physik, Lehrstuhl IV: Weltraum- und Astrophysik, Ruhr-Universität Bochum D-44780 Bochum, Germany\\
\email{theo.joubaud@cea.fr} \\
\email{isabelle.grenier@cea.fr}
}

\date{Received <date> /
Accepted <date>}

\abstract {}
{The nearby Orion-Eridanus superbubble, which was blown by multiple supernovae several million years ago, has likely produced cosmic rays. Its turbulent medium is still energised by massive stellar winds and it can impact cosmic-ray transport locally. The \g radiation produced in interactions between cosmic rays and interstellar gas can be used to compare the cosmic-ray spectrum in the superbubble and in other regions near the Sun. It can reveal spectral changes induced in GeV to TeV cosmic rays by the past and present stellar activity in the superbubble.}
{We used ten years of data from the \Fermi Large Area Telescope (LAT) in the 0.25-63 GeV energy range to study the closer (Eridanus) end of the superbubble at low Galactic latitudes. We modelled the spatial and spectral distributions of the \g rays produced in the different gas phases (atomic, molecular, dark, and ionised) of the clouds found in this direction. The model included other non-gaseous components to match the data.}
{We found that the \g-ray emissivity spectrum of the gas along the outer rim and in a shell inside the superbubble is consistent with the average spectrum measured in the solar neighbourhood. It is also consistent with the cosmic-ray spectrum directly measured in the Solar System. This homogeneity calls for a detailed assessment of the recent supernova rate and current census of massive stellar winds in the superbubble in order to estimate the epoch and rate of cosmic-ray production and to constrain the transport conditions that can lead to such homogeneity and little re-acceleration.
We also found significant evidence that a diffuse atomic cloud lying outside the superbubble, at a height of 200-250~pc below the Galactic plane, is pervaded by a 34\% lower cosmic-ray flux, but with the same particle energy distribution as the local one. Super-GeV cosmic rays should freely cross such a light and diffuse cirrus cloud without significant loss or spectral distorsion. We tentatively propose that the cosmic-ray loss relates to the orientation of the magnetic field lines threading the cirrus, which point towards the halo according to the dust polarisation data from Planck. Finally, we gathered the present emissivity measurements with previous estimates obtained around the Sun to show how the local cosmic-ray flux decreases with Galactic height and to compare this trend with model predictions. }
{}

\maketitle

\section{Introduction}
Superbubbles form around starburst regions under the combined and sustained influence of the ionising radiation and energetic winds of massive stars plus a series of supernovae (SNe). They have long been considered as potential sites for cosmic-ray (CR) production because of the collective action of the multiple shock waves they contain \citep{Montmerle79,Cesarsky83,Bykov01Toptygin,Bykov01,Parizot04,Ferrand10}. The abundances of CR nuclei also require a mix of massive-star outflows, supernova ejecta, and interstellar matter that is easily found in superbubbles \citep[][and references therein]{Lingenfelter18}. The relative fractions in the mix depend on the uncertain stellar yields for heavy nuclei \citep{Binns19}. Superbubbles can also alter the distribution of Galactic CR diffusing through them because of the broad spectrum of magnetic turbulence powered by their multi-scale flows.

Modelling the acceleration, transport, and losses of CR inside the turbulent and multiphasic medium of a superbubble is particularly complex \citep[see the review by][]{Bykov14}. Intermittent acceleration by individual shocks, followed by interactions with large-scale compression and rarefaction waves, can produce hard CR spectra in the MeV-TeV range 
\citep{Bykov01}. The momentum distribution of the particles evolves over a few million years (Myr). It asymptotically tends toward a power law for $p^2 f(p) \propto p^{-\gamma}$, with a \g index ranging between two and three depending on the particle injection momentum: $\gamma=3$ for the acceleration of suprathermal particles and $\gamma=2$ for the re-acceleration of relativistic CR. Such superbubbles can transfer 10\% or more of their kinetic power to CR \citep{Bykov01}. 

Superbubbles can also affect Galactic CR entering them. Particle interactions with magnetohydrodynamic (MHD) waves in the turbulent plasma can re-accelerate the CR despite pion losses along their path. For efficient superbubbles, with acceleration time scales much shorter than the diffusive escape time scale from the bubble, the resulting hadronic \g-ray emission can exhibit power-law differential spectra as hard as $E_{\gamma}^{-0.3}$ in the 1-100 GeV energy band \citep{Tolksdorf19}. The re-accelerated CR should accumulate at the periphery of the superbubble as the enhanced turbulence prevents their penetration deep into the interior \citep{Tolksdorf19}. Conversely, CR that are produced internally can remain confined in the superbubble medium for typically 300 kyr for 100~GeV particles in a 25-50~pc superbubble if the diffusion lengths are 100 times shorter than in the standard interstellar medium (ISM) \citep{Ackermann11}. Superbubbles have therefore enough energy and magnetic turbulence to substantially modify the CR properties, but whether they have a positive (acceleration) or negative (confinement and losses) feedback is still an open question that needs clarifications from observations. 

\begin{figure}
\resizebox{\hsize}{!}{\includegraphics{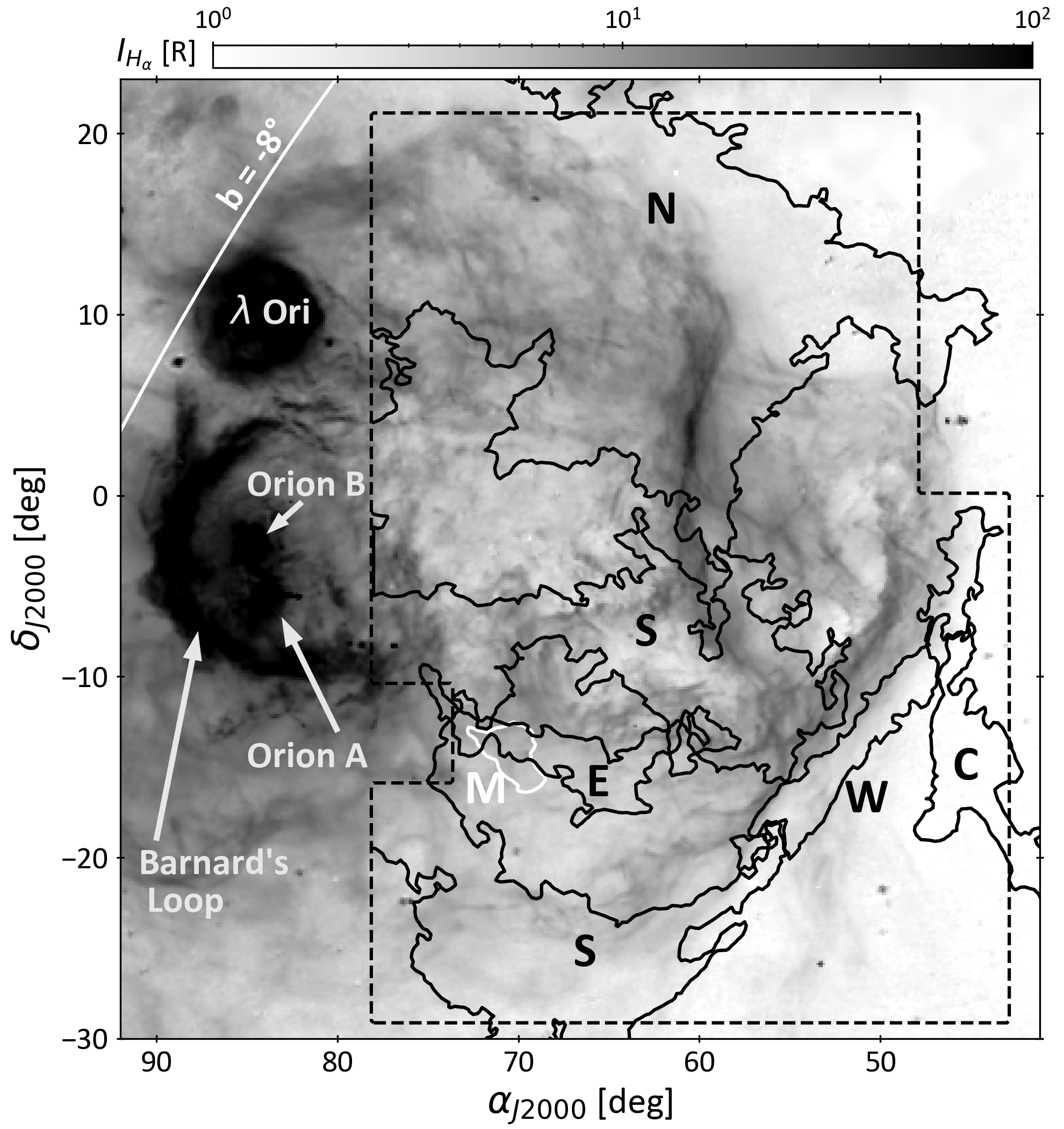}} 
\caption{\halpha intensity map of the Orion-Eridanus superbubble based on VTSS, SHASSA and WHAM data. The dashed line traces the perimeter of the present analysis. The white labels on the left hand side mark the key \halpha features towards Orion. In the analysis region, the black contours delineate the main \hi shells related to the superbubble, i.e the \northrim (N), \southloop (S), \eastshell (E), and \westrim (W), and the Eridu cirrus (C) lying outside the superbubble \citep{Joubaud19}. The molecular MBM 20 cloud in front of the superbubble edge is outlined in white. The line of constant Galactic latitude b=-8\degr\ in the upper left corner indicates the orientation of the Galactic plane.}
\label{fig:Halpha}
\end{figure}

In 2011, \citeauthor{Ackermann11} reported an extended excess of hard \g-ray emission from the Cygnus X starburst region, which they interpreted as a cocoon of freshly accelerated CR inside the 60~pc-wide superbubble. The latter has been carved by the stellar winds and ionisation fronts from young stellar clusters including Cyg OB2 and NGC 6910. Cyg OB2 is one of the most massive clusters in the Galaxy at a distance of $1755^{+23}_{-19}(stat)^{+373}_{-261}(syst)$~kpc \citep{Berlanas19} and the NGC 6910 cluster is part of the Cyg OB9 association at a distance of $1600 \pm 200$~pc \citep{Kolaczkowski04}.
The $E_{\gamma}^{-2.1}$ differential \g-ray spectrum extends to 100 GeV in the \Fermi LAT data and softens to $E_{\gamma}^{-2.6}$ in the 0.4-10 TeV range in the MILAGRO, ARGO-YBJ, and HAWC data \citep{Bartoli14,Hona17}. The apparent break energy around 1 TeV is uncertain because of the confusion with other extended TeV sources in these crowded directions. The \g-ray luminosity does not exceed a few percent of the wind power in the stellar clusters. Star formation has progressed across the superbubble from the low longitude end $\gtrsim 10$ Myr ago towards the younger Cyg OB2 cluster \citep{Berlanas18}. The latter has formed stars more or less continuously between 1 and 7 Myr ago and its mass distribution suggests that its most massive members have already evolved to their supernova end state \citep{Wright2015}. Five or six of them may have exploded within the last 1-2 Myr \citep{Lingenfelter18}. NGC 6910 has a comparable age of 
$6\pm 2$ Myr \citep{Kolaczkowski04}, but is also part of the older and broader Cyg OB9 association. 

A second possible cocoon of young CR has been reported in \g rays further away in the Galactic disc, around l=25\degr\ and b=0\degr, in a gas cavity found near a candidate massive OB association, G25.18+0.26, seen in X rays \citep{Katsuta17}. The distance estimate of 7.7~kpc implies a three times larger bubble (210x170~pc) and a 14 times larger \g-ray luminosity than for the Cygnus X cocoon, but with a similarly hard emission spectrum extending to 100 GeV. The large distance and heavy extinction, however, hamper further optical analyses of the stellar cluster. Confusion with other \g-ray sources and with the bright Galactic emission further hampers \g-ray analyses of high-energy CR until better resolved observations with the Cherenkov telescope array (CTA) become available  \citep{Acharya13}. Both the Cygnus X and G25.0+0.0 cocoons unfortunately lie in confused directions respectively tangent to the Local spiral arm and Scutum-Centaurus arm.

The Orion-Eridanus superbubble is the nearest example of a superbubble. This over-pressured and elongated cavity, about 200~pc in width and 250~pc in length, slowly expands at a velocity of 20 \kmpers \citep[][alias \citetalias{Joubaud19} hereafter]{Joubaud19}. Its close end, in the Eridanus constellation, approaches the Local Bubble at a distance of 150-200~pc from the Sun. Its far end lies slightly beyond the 388 $\pm$ 5~pc distant, 1-2 Myr old stellar clusters of the Orion Nebula \citep{Pon16,Kounkel17,Zari17}. It has been formed by the ionising radiation and energetic winds of tens of massive stars and by a series of 10 to 20 supernovae \citep{Bally2008} that have occurred at a rate of about 1 Myr$^{-1}$ over the past 12 Myr \citep{Voss10}. The composite structure of the superbubble has likely evolved in space and time from near to far, along the blue stream of massive stars identified by \citet{Pellizza05} and \citet{Bouy15} in front of the Orion clouds. The stream extends over a length of 150~pc along the major axis of the superbubble. The stellar ages span from 20 Myr (near) down to 1 Myr (far). 

The near end of the cavity is filled with a mix of hot and warm plasmas with temperatures of order (3-9) and $(0.3-1.2)$~MK, and gas densities of order 0.005 and 0.05 cm$^{-3}$, respectively \citepalias{Joubaud19}. It is bounded by shells of neutral gas that have been swept-up and compressed by the expanding outer shock wave. Several shells have been isolated in position, velocity, and distance \citepalias{Joubaud19} and they were used here to probe the CR flux in various parts of the superbubble. They are sketched against the \halpha intensity map of the superbubble in Fig. \ref{fig:Halpha} and their \hi column density and CO intensity maps are presented in Fig. 2 of \cite{Joubaud19}. 

We studied the near part of the superbubble using \g-ray data obtained with the \Fermi LAT \citep{Atwood2009}. This part lies at large (negative) Galactic latitudes, from -50 to -15\degr, away from Galactic confusion. To do so, we modelled the \g-ray emission coming from the hadronic interactions of CR with the different gas phases and clouds of the superbubble. We coupled \g-ray and dust analyses to trace the total gas, in particular the dark neutral medium which is not detected via \hi and CO observations. 
Our analysis yielded \g-ray emissivities per gas nucleon in the atomic phase of the different clouds, where the \hi line data allow a direct estimation of the gas mass pervaded by the CR. We could thus compare the CR flux in the superbubble with the average measured in the local ISM and in the Solar System \citep{Casandjian15}. Gamma-ray emissivity estimates in the local ISM have indeed shown that the CR flux is quite uniform within a few hundred parsecs from the Sun and that it is in agreement with the direct measurements obtained in the Solar System \citep[see the review by][]{Grenier15}. Other works, based on different interaction cross-sections \citep{Strong15, Orlando18}, found a $\sim$30\% discrepancy in the emissivities between direct measurements in the Solar System and the local ISM average. Residual cloud to cloud variations of order 10\% are commensurate with systematic uncertainties in the atomic gas mass that has served for the measurements. 

The \g-ray emissivities per gas nucleon have also been shown to decrease with increasing altitude above the Galactic plane, by about  50\% within 2~kpc \citep{Tibaldo15}. Our analyses allowed the study of the CR flux in a diffuse cirrus cloud lying in the vicinity of the superbubble, below it with respect to the Galactic plane.

The paper is structured as follows : data are presented in Sect. \ref{sec:data}, we describe the \g-ray and dust models in Sect. \ref{sec:Analysis}. The results are presented in Sect. \ref{sec:results} for the superbubble first and then the cirrus. The results are discussed in Sect \ref{sec:Discussion} in the same order.

\section{Data}
\label{sec:data}

The present analysis is based on the same gas data and on the same atomic and molecular cloud separation as in the study of the dynamics and gas content of the superbubble presented in \citetalias{Joubaud19}. We used the same analysis region towards the Eridanus part of the superbubble, extending in equatorial coordinates from 43\degr\ to 78\degr\ in right ascension and from -29\degr\ to 21\degr\ in declination, as shown in Fig. \ref{fig:Halpha}. We masked out two 5\degr-wide areas on the western and eastern sides of the region to avoid complex gas distributions in the background. All maps were projected onto the same 0\fdg25-spaced Cartesian grid, that is appropriate for the \g-ray counts observed at very high Galactic latitudes.

	\subsection{Gamma-ray data}
	\label{sec:gammadata}
We used ten years of Pass 8 photon data provided by the LAT between 0.25 and 63 GeV \citep{Atwood13, Bruel18}. This energy range was chosen to preserve an angular resolution better than approximately 2\degr\ 68\% containment angle at low energy and to have sufficient photon statistics at high energy. We used the associated instrument response functions P8R3\_SOURCE\_V2 (for the various PSF event types), and the corresponding isotropic spectrum for the extragalactic and residual instrumental backgrounds\footnote{https://fermi.gsfc.nasa.gov/ssc/data/access/lat/BackgroundModels.html}. Tight selection criteria were used: SOURCE class selection, PSF 1, 2 and 3 event types, photon arrival directions within 100\degr\ or 105\degr\ of the Earth zenith depending on the photon energy and PSF type (100\degr\ below 0.6~GeV for PSF 1-2 and below 0.4~GeV for PSF 3, 105\degr\ otherwise). Such criteria reduce the contamination by residual cosmic rays and by Earth atmospheric \g rays in the photon data. The instrument functions, the exposure map, the \g-ray emissivity spectrum of the local interstellar gas, \qlis \citep{Casandjian15}, and the spectrum of the isotropic background were evaluated in 12 energy bins, 0.2 dex in width and centred from $10^{2.5}$ to $10^{4.7}$ MeV. We took the energy resolution of the LAT into account when modelling the data.
To ensure photon statistics robust enough to follow details in the spatial distributions of the different interstellar components, we analysed the data in seven broad and independent energy bands, bounded by $10^{2.4}$, $10^{2.6}$, $10^{2.8}$, $10^{3.0}$,  $10^{3.2}$, $10^{3.6}$, $10^{4.0}$, and $10^{4.8}$ MeV. 

	\subsection{\hi and CO data, as well as cloud separation}
In order to trace the atomic gas, we used the 16\farcm2 resolution HI4PI survey \citep{HI4PI}, with a velocity resolution of 1.49 km/s in the local standard of rest (LSR). We selected velocities between -90 and +50 km s$^{-1}$ to exclude the \hi emission coming from the high-velocity clouds that lie in the hot Galactic corona, far behind the local medium we are interested in \citep{Wakker2008}. We checked that these high-velocity clouds were not detected in \g rays and we removed them from our model.

In order to trace the molecular gas, we used the 8\farcm5 resolution $^{12}$CO (J=1$-$0) observations at 115 GHz from the moment-masked CfA CO survey \citep{Dame2001, Dame2004}. We completed this dataset with the CO observations of the MBM 20 cloud that were obtained with the Swedish-ESO Submillimetre Telescope (SEST). They were kindly provided by D. Russeil \citep{Russeil2003}.

We decomposed the \hi and CO velocity spectra into individual lines and we used this information to identify and separate eight nearby cloud complexes that are coherent in position, velocity, and distance. Details of the method are presented in \citetalias{Joubaud19}. The resulting maps were projected onto a 0\fdg25-spaced Cartesian grid in equatorial coordinates. The main entities associated with the superbubble are the \northrim, \westrim, \southloop along the outer rim, and the \eastshell in the interior. We have also identified the Eridu\footnote{named after the Babylonian constellation `Star of Eridu' possibly at the origin of the name of the Eridanus Greek constellation} atomic cirrus cloud that lies outside the superbubble, at a comparable distance, but further away from the Galactic plane. The compact MBM 20 molecular cloud lies just in front of the superbubble edge, between the Local Bubble and the Orion-Eridanus superbubble. The relative positions of all these clouds in the sky are displayed in Fig. \ref{fig:Halpha}. The analysis region partially intercepts other cloud complexes, the edge of the nearby Taurus cloud, and part of the Cetus and North Taurus complexes which are located behind the superbubble \citep{Remy17}. Their contributions to the \g-ray emission were taken into account in the model, as well as the faint atomic background from the Galactic disc. 

\subsection{Dust}
\label{sec:dataDust}
In order to trace the dust column density, we used the optical depth map obtained at 353 GHz, $\tau_{353}$, by \citet{PlanckXVII14}. It was obtained by modelling the dust thermal emission recorded by \Planck and the Infrared Astronomical Satellite (\IRAS) with a modified black body, $I_{\nu} {=} \tau_{353\,{\rm GHz}} B_{\nu}(T_d)(\nu/353\,{\rm GHz})^{\beta}$, with $T_d$ the dust colour temperature and $\beta$ the spectral index. We degraded the optical-depth map from its original resolution of 5\arcmin\ down to 16.2\arcmin\ to match the resolution of the \hi data and we projected it onto a 0\fdg25-spaced Cartesian grid. The dust opacity, or optical depth per gas nucleon, $\sigma_{\nu}$, is defined according to $\tau_{\nu} = \frac{I_{\nu}}{B_{\nu}(T)} = \sigma_{\nu} N_H$.

\subsection{Ionised gas}
	\label{sec:HaData}
Ionised gas is visible in \halpha emission. It is displayed in Fig. \ref{fig:Halpha} using the data of \citet{Finkbeiner2003}, which is a composite map of the Virginia Tech Spectral line Survey (VTSS), the Southern H-Alpha Sky Survey Atlas (SHASSA) and the Wisconsin H-Alpha Mapper (WHAM). 
With visual extinctions ranging from 0.3 to 2.2 across the field, the \halpha emission is partially absorbed, in particular towards the \northrim clouds. Moreover, \halpha emission traces only the recombining gas and it does not linearly scale with the column density of ionised gas, \nhii. Its intensity is proportional to the emission measure : $I_{\halpha} \propto \int n_i n_e ds$, with $n_i$ the ion density, $n_e$ the electron density and the integration is along the line of sight. Free-free emission at mm wavelengths can also trace the ionised gas, but the  most recent estimates inferred from the Planck data are still too heavily contaminated by dust emission in our analysis region. We therefore created a uniform template in order to keep the morphology of the main \halpha arcs. We set the template value to one within the 14~R contour of \halpha emission and zero outside. The normalisation was left free in our fits (see Sec. \ref{sec:Analysis}). We note this template $K_{\halpha}$.

The hot ionised gas that fills the superbubble unfortunately yields column densities below $5\times 10^{19}$~\persqcm that are too low to be detected in \g rays with the LAT. We thus did not include this gas in the model.

\begin{figure}
\centering
\resizebox{0.9\hsize}{!}{\includegraphics{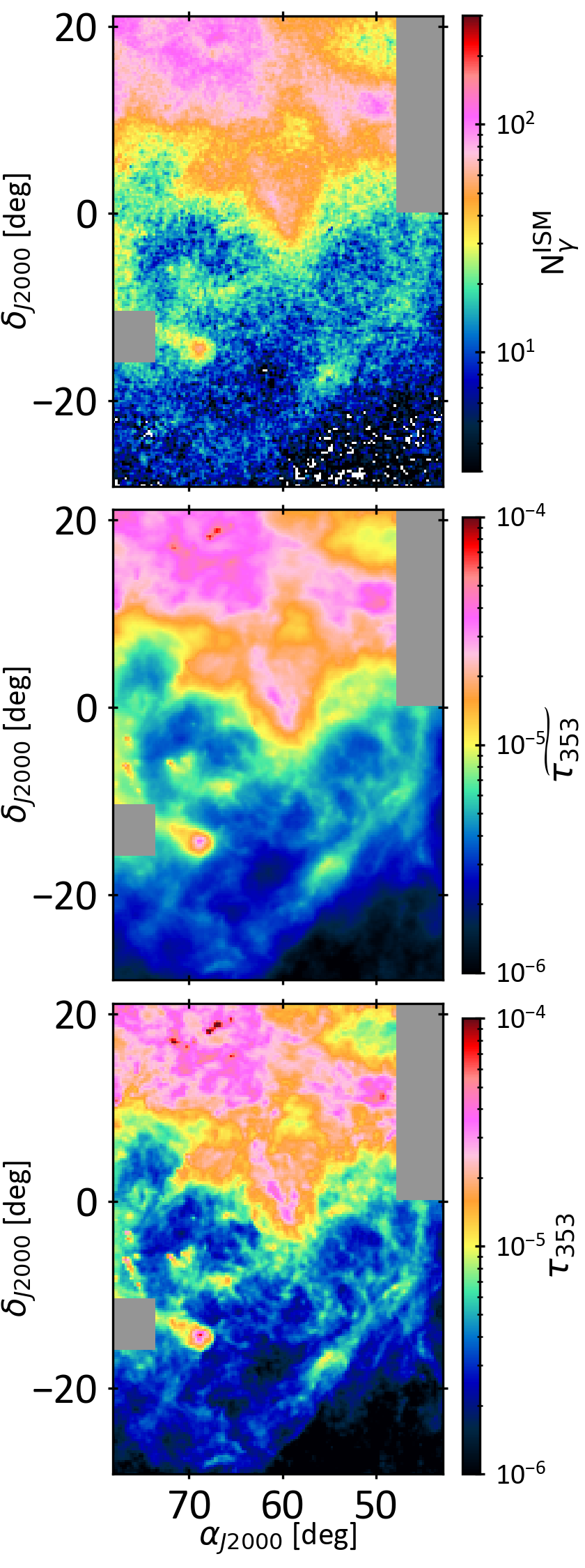}}
\caption{\textit{Top:} \g-ray counts of gaseous origin recorded in the 0.25-63~GeV band in a 0\fdg25 pixel grid. \g-ray emissions other than due to CR interactions with the gas were subtracted. \textit{Middle:} dust optical depth measured at 353~GHz and displayed at the \Fermi-LAT angular resolution for comparison. \textit{Bottom:} dust optical depth measured at 353 GHz at its original resolution of 5\arcmin.}
\label{fig:Dustcol}
\end{figure}

\section{Models and analyses}
\label{sec:Analysis}

The intensity of the hadronic \g radiation corresponds to the integral along the lines of sight of the CR flux times the gas volume density. The LAT data we have used probe CR with energies above a few GeV that uniformly permeate all gas phases up to the molecular phase seen in CO at 115 GHz \citep[][and references therein]{Grenier15,Remy17}. The \g-ray intensity therefore scales with the total gas column density, \nh, and can be modelled by a linear combination of the column densities present in the ionised, atomic, molecular, and dark neutral phases. The average \g-ray emissivity per gas nucleon in a cloud can only be inferred in the \hi phase where we can directly measure the hydrogen column densities, provided corrections for self absorption of the \hi lines. This is not possible in the other gas phases where the knowledge of the CO-to-\hd conversion factors and of the dust emission opacities is required to infer the gas mass. These factors are, moreover, known to vary from cloud to cloud and with environmental conditions \citep{Remy17}. To model the total \g-ray emission, additional ancillary components needed to be considered like the Galactic inverse Compton (IC) radiation, point sources, the isotropic emission accounting for extragalactic and instrumental backgrounds, and the emissions from the Sun and the Moon.

    \subsection{The dark neutral medium (DNM)}
The dark neutral phase contains large amounts of gas at the \hi-\hd interface, but it is invisible in \hi and CO line emission because of \hi self-absorption and because of the photo-dissociation and weak excitation of CO molecules in diffuse \hd. In order to trace the DNM gas, we iteratively coupled the \g-ray and dust analyses as dust column densities also trace the total \nh in the local ISM. This is illustrated in Fig. \ref{fig:Dustcol} where we convolved the dust optical depth map with the LAT PSF on the one hand, and, on the other hand, we subtracted from the \g-ray data the non-gaseous emissions obtained from our best-fit model (Sec. \ref{sec:GammaModel}). Figure \ref{fig:Dustcol} shows strong similarities in the spatial distributions of the dust and \g-ray gas tracers, but it also reveals differences in their dynamical range in several places. It has been shown in particular that the dust optical depth does not linearly trace the total gas column density in the dense molecular phase ($\nh > 3 \times 10^{21}$~\persqcm) because the dust grains evolve and their emission properties change \citep{PlanckXVII14,PlanckXXVIII15,Remy17}. The DNM phase for which we use dust data in our analysis is, however, diffuse enough for the linear approximation to hold. 
We therefore extracted from the dust and interstellar \g-ray maps the significant signals found above the \hi and CO expectations and we used the spatial correlation between these signals to infer the additional gas column densities that correspond to the DNM at the \hi-\hd transition.

	\subsection{Gamma-ray model}
	\label{sec:GammaModel}
Earlier studies indicated that the bulk of the Galactic CR radiating at 0.1–100 GeV have diffusion lengths much larger than typical cloud dimensions and that they permeate all the \hi-bright, DNM, and CO-bright gas phases. The observed interstellar \g-ray emission can therefore be modelled, to first order, by a linear combination of the contributions of the different clouds seen along the lines of sight, including their different gas phases (\halpha, \hi, DNM, CO). In order to convolve the gas maps with the LAT response functions, we assumed that the \g-ray emissivity spectrum of each gas phase of each cloud follows the average one obtained in the local ISM \citep[\qlis(E), ][]{Casandjian15}. We left, however, a free normalisation factor for the emissivity of each component in each energy band to account for possible deviations in CR density and spectrum.

The model includes other radiation components such as the Galactic IC intensity, I$_{\rm{IC}}(\alpha,\delta$, E), the isotropic intensity mentioned above, I$_{\rm{iso}}$(E), intensities from the Sun and the Moon, I$_{\rm{SM}}(\alpha,\delta$, E), and point sources with individual flux spectra S$_j$(E). The \g-ray intensity I($\alpha,\delta$, E), expressed in cm$^{-2}$ s$^{-1}$ sr$^{-1}$ MeV$^{-1}$, can thus be modelled in each energy band as :
\begin{equation}
  \label{eq:gam}
  \begin{split}
I(\alpha,\delta,E)\; &=\; \qlis(E) \times \Bigg[ \sum\limits_{i=1}^{9} q_{\hi,i}(E) \, N_{\hi,i}(\alpha,\delta) \\
            &+ \sum\limits_{i=1}^{3} q_{\rm{CO},i}(E) \, W_{\rm{CO},i}(\alpha,\delta) + q_{\halpha}(E) \, K_{\halpha}(\alpha,\delta) \\
            &+ q_{\rm{DNM}}(E) \, \tau_{353}^{\rm{DNM}}(\alpha,\delta) \Bigg] + q_{\rm{iso}}(E) \, I_{\rm{iso}}(E) \\
            &+ q_{\rm{IC}}(E) \, I_{\rm{IC}}(\alpha,\delta,E) + q_{\rm{SM}}(E) \, I_{\rm{SM}}(\alpha,\delta,E)\\
			&+ \sum\limits_{j} q_{S_j}(E) \,  S_{j}(E) \, \delta(\alpha-\alpha_j,\delta-\delta_j)\\
			&+ q_{S_{\rm{ext}}}(E) \, S_{\rm{ext}}(\alpha,\delta,E)
    \end{split}
\end{equation}
where $N_{\hi,i}(\alpha,\delta)$ denotes the \hi column density map  of each component, $W_{\rm{CO},i}(\alpha,\delta)$ gives the CO line integral intensity map, and $K_{\halpha}(\alpha,\delta)$ is the \halpha template described in Sec. \ref{sec:HaData}. $\tau_{353}^{DNM}$ stands for the DNM dust map resulting from the joint dust and \g-ray analyses (further detailed in Sect. \ref{sec:Iteration}). To account for the spill-over of emission produced outside the analysis region, but reconstructed inside it, we modelled point sources and interstellar contributions in a region 6\degr\ wider than the analysis region.

\begin{figure*}
	\centering
	\includegraphics[width=17cm]{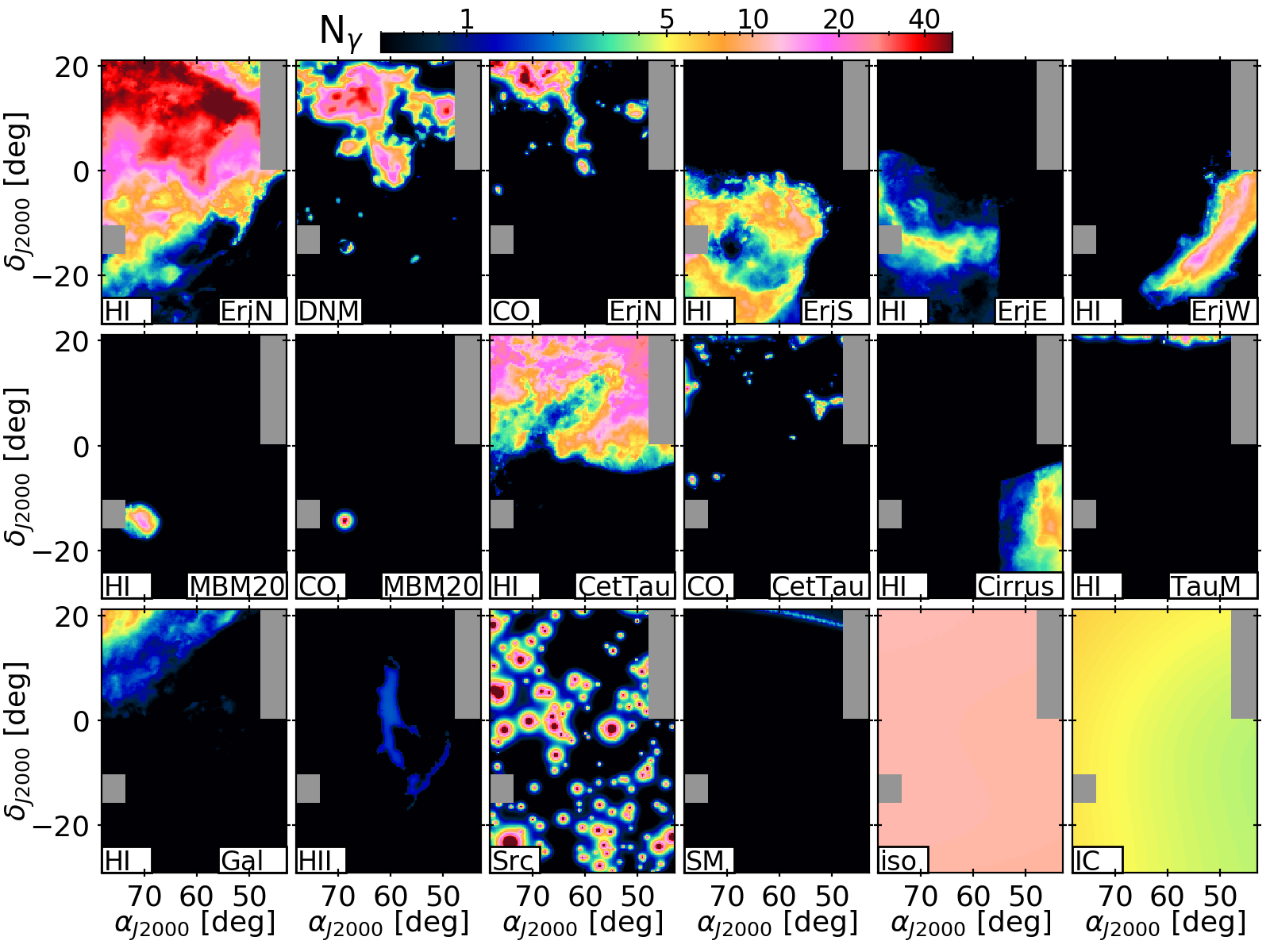}
	\caption{Photon yields arising in the \g-ray model for the best-fit to the data in the 0.25-63 GeV band on a 0\fdg25 grid, from the \hi and CO phases of the \northrim (EriN), \southloop (EriS), \eastshell (EriE), \westrim (EriW), MBM 20, Cetus and Taurus, Eridu cirrus, Main Taurus and Galactic disc clouds, from the DNM gas column density, from the \halpha template (HII), from the \g-ray point sources, from the Sun and Moon emissions (SM), from the isotropic background (iso) and from the Galactic IC emission.}
	\label{fig:GamModel}
\end{figure*}

All the components are illustrated in Fig. \ref{fig:GamModel}. Three clouds have atomic envelopes and molecular cores (the superbubble \northrim, the foreground MBM 20 cloud, and the background Cetus-North Taurus clouds). The other five clouds (the superbubble \southloop, \westrim, and \eastshell, the Eridu cirrus, and the edge of the foreground Taurus cloud) as well as the Galactic disc background are only detected in \hi. The DNM map is dominated by dark gas in the \northrim and a small contribution from MBM 20. 

The input \qlis spectrum was based on four years of LAT data and on the correlation between the \g radiation and the \nhi column densities derived from the LAB survey, for a spin temperature of 140 K, at latitudes between 7\degr\ and 70\degr\ \citep{Casandjian15}. The $q_{\hi,i}$(E) normalisation factors in the model can therefore compensate for cloud-to-cloud variations in CR flux or spectrum. For each cloud, the final average \g-ray emissivity spectrum per gas nucleon in the atomic phase is the product of the \qlis(E) spectrum and of the best-fit $q_{\hi,i}$(E) normalisation in each energy band.

The model includes 151 points sources from the \Fermi LAT 8-year FL8Y source list\footnote{\url{https://fermi.gsfc.nasa.gov/ssc/data/access/lat/fl8y/}} inside the analysis region. Their flux spectra, $S_{j}(E)$, were computed with the spectral characteristics provided in the catalogue. Their $q_{S_j}(E)$ normalisation in the model allows for possible changes due to the longer exposure used here and to the use of a different interstellar background for source detection in FL8Y. The sources in the analysis region were fitted individually. The sources present in the 6\degr-wide peripheral band around the analysis region were merged into a single map, $S_{\rm{ext}}(\alpha,\delta,E)$, and its normalisation, $q_{S_{\rm{ext}}}$(E), was left free in each energy band. We checked that its best-fit normalisation was compatible with one.

The isotropic emission spectrum was determined with the FL8Y interstellar background model\footnote{\url{https://fermi.gsfc.nasa.gov/ssc/data/access/lat/BackgroundModels.html}}. The \g-ray emission observed in the region also includes a contribution from the large-scale Galactic IC emission emanating from the interactions of CR electrons with the Galactic interstellar radiation field. The GALPROP\footnote{\url{http://galprop.stanford.edu}} parameter file 54-LRYusifovXCO4z6R30-Ts150-mag2 was used to generate an energy-dependent template of the Galactic IC emission across the analysis region \citep{Ackermann12}. The Sun and the Moon paths also cross this region, so we took their \g-ray emissions into account. This was done following the method described by \citet{Johannesson13}, reproducing the interactions of CR electrons with the solar radiation field and the hadronic interactions of CR nuclei with the Sun and the Moon.
The Galactic IC emission model, the isotropic emission spectrum, and the Sun and Moon emission models were derived from earlier LAT data. Therefore, we left their respective scalings free in each energy band to allow for possible changes due to the longer exposure used here and to the use of a different interstellar background.

In order to compare the modelled photon map predictions with the LAT photon data in the different energy bands, we multiplied each component map by the energy-dependent exposure and we convolved it with the energy-dependent PSF. We also convolved the spectra in each pixel with the LAT energy resolution. The convolution of the gas maps with the response functions was done for each PSF event type independently in small energy bands. We then summed the photon count maps in energy and for the different PSF types to get the modelled count maps in the seven energy bands of the analysis. Figure \ref{fig:GamModel} gives the photon yields that were obtained in the entire energy band from the best fit. It shows that the emission originating from the gas dominates over other types of emissions and that the LAT angular resolution makes it possible to separate the various clouds, as well as the different gas phases within the clouds, based on their spatial distribution. The $q$ model parameters were fitted to the LAT photon data in each energy band using a binned maximum-likelihood with Poisson statistics.

	\subsection{Dust model}
In the case of a uniform dust-to-gas mass ratio and uniform mass emission coefficient of the grains, the dust optical depth linearly scales with the total \nh. We therefore modelled $\tau_{353}(\alpha,\delta)$ in each direction as a linear combination of the same gaseous contributions as in the \g-ray model, with free normalisations to be fitted to the data, as in \citet{PlanckXXVIII15, Remy17}. We added a free isotropic term, y$_{iso}$, to account for the residual noise and the uncertainty in the zero level of the dust data \citep{PlanckXI14}. The $\tau_{353}(\alpha,\delta)$ model can be expressed as
\begin{equation}
    \label{eq:dust}
    \begin{split}
        \tau_{353}(\alpha,\delta)  = \sum\limits_{i=1}^{9} y_{\hi,i} \, N_{\hi,i}(\alpha,\delta) + \sum\limits_{i=1}^{3} y_{\rm{CO},i} \,  W_{\rm{CO},i}(\alpha,\delta)
        \\ + \ydnm \, \nhdnm(\alpha,\delta) + y_{\halpha} \,  K_{\halpha}(\alpha,\delta) + y_{iso},
    \end{split}
\end{equation}
where \nhdnm$(\alpha,\delta)$ stands for the DNM gas column density map iteratively built from the coupled analyses of the \g-ray and dust data (further detailed in Sect. \ref{sec:Iteration}). 

The $y$ model parameters were estimated using a $\chi^2$ minimisation. We expect the model uncertainties to exceed the measurement errors in $\tau_{353}(\alpha,\delta)$ because of potential variations in grain properties through the clouds and because of the limitations of the gas tracers (survey sensitivities, emission saturation, self-absorption, etc.). As we cannot precisely determine the model uncertainties, we set them to a fractional value of the model and we determined this fraction to be 18\% by reaching a reduced $\chi^2$ of unity. This fraction is larger than the 3\% to 9\% uncertainties in the $\tau_{353}$ values measured across this region \citep{PlanckXI14}.

	\subsection{Analysis iteration}
	\label{sec:Iteration}
In order to extract the DNM gas present in both the dust and \g-ray data in addition to the gas seen in \hi and CO emission, we iteratively coupled the \g-ray and dust models. We built residual maps between the data (dust or \g rays) and the best-fit contributions from the \nhii, \nhi, \wco and other non-gaseous components. We kept only the positive residuals with high significance above the noise (a simple cut at zero would induce an offset bias in the next model). To do so, we de-noised the residual maps using the multi-resolution support method implemented in the MR filter software \citep{Starck98}. We used a multi-resolution thresholding filter, with six scales, Gaussian noise and a 2-$\sigma$ threshold to filter the dust residuals. For the \g-ray ones, we used seven scales, Gaussian and Poisson noise, and a 3-$\sigma$ threshold. The DNM template estimated from the dust was provided to the \g-ray model ($ \tau_{353}^{\rm{DNM}}$ in equation \ref{eq:gam}); conversely, the DNM column density map derived from the \g rays was provided to the dust model (\nhdnm in equation \ref{eq:dust}). We started the iterative process by fitting the dust optical depth with the \hi, CO, \halpha and isotropic components to build the first DNM map for the \g-ray model. We then iterated between the \g-ray and dust models until reaching a saturation in the log-likelihood value of the fit to the \g-ray data (from the third to the fourth iteration). We checked that the resulting DNM maps were consistent in morphology and column density with the maps obtained by \citet{Remy17} in the intersection region between the two analyses.

At each iteration, the estimates of the $q$ and $y$ model coefficients as well as the DNM maps changed, and the likelihood significantly improved since there was less and less need for the other components, in particular the \hi and CO ones, to compensate for the missing gas. They still do at some level because the DNM template provided by the \g rays or dust optical depth have limitations (e.g., dust opacity variations, limited \g-ray sensitivity).

\section{Results}
\label{sec:results}

Before presenting in Sect. \ref{sec:qSB} and \ref{sec:qEridu} the \g-ray emissivity spectra obtained in the different clouds from the best-fit models described in Equations \ref{eq:dust} and \ref{eq:gam}, we discuss how the fits to the \g-ray and dust data improved with optical depth corrections to the atomic \nhi column densities (Sect. \ref{sec:Ts}) and with the addition of gas templates other than \hi and CO (Sect. \ref{sec:DNMdet} and \ref{sec:HIIdet}). We also assess the robustness of the best-fit models with jackknife tests in Sect. \ref{sec:jack} and we present residual maps in Sect. \ref{sec:residual}.


	\subsection{\hi optical depth correction}
	\label{sec:Ts}
In the range of \nhi column densities studied here, \citet{Nguyen19} showed that a simple isothermal correction of the \hi emission spectra, with a uniform spin temperature (\tspin) across the cloud, provides better than 10\% estimates of the more precise \nhi values inferred from the combination of emission and absorption \hi spectra. Other correction methods are less efficient. In order to account for the unknown level of \hi opacity in the different \nhi maps, we have repeated our dust and \g-ray analyses for ten spin temperatures (100, 125, 150, 200, 300, 400, 500, 600, 700, 800~K) and for the optically thin case which yields the minimum amount of gas. We applied the same spin temperature to all the clouds as we could not test all the possible combinations of individual spin temperatures. The best-fit residual maps show that a single spin temperature can reproduce the \g-ray data without any significantly deviant cloud. 

The \g rays can help constrain the average level of \hi optical-depth correction applicable to the whole region by comparing the \tspin-dependent contrast of the \nhi maps with the spatial structure of the \g-ray flux emerging from the \hi gas. We found that the maximum log-likelihood value of the fits significantly preferred a spin temperature of 100~K in this region (see appendix \ref{ap:Tspin}). We could not test spin temperatures below 100~K in our isothermal approach as its value needs to be above the highest values of the brightness temperature, which are close to 100~K in our region. In the following, we present the results obtained for this temperature of 100~K, unless otherwise mentioned.

 \begin{figure}
	\begin{center}
	\resizebox{\hsize}{!}{\includegraphics{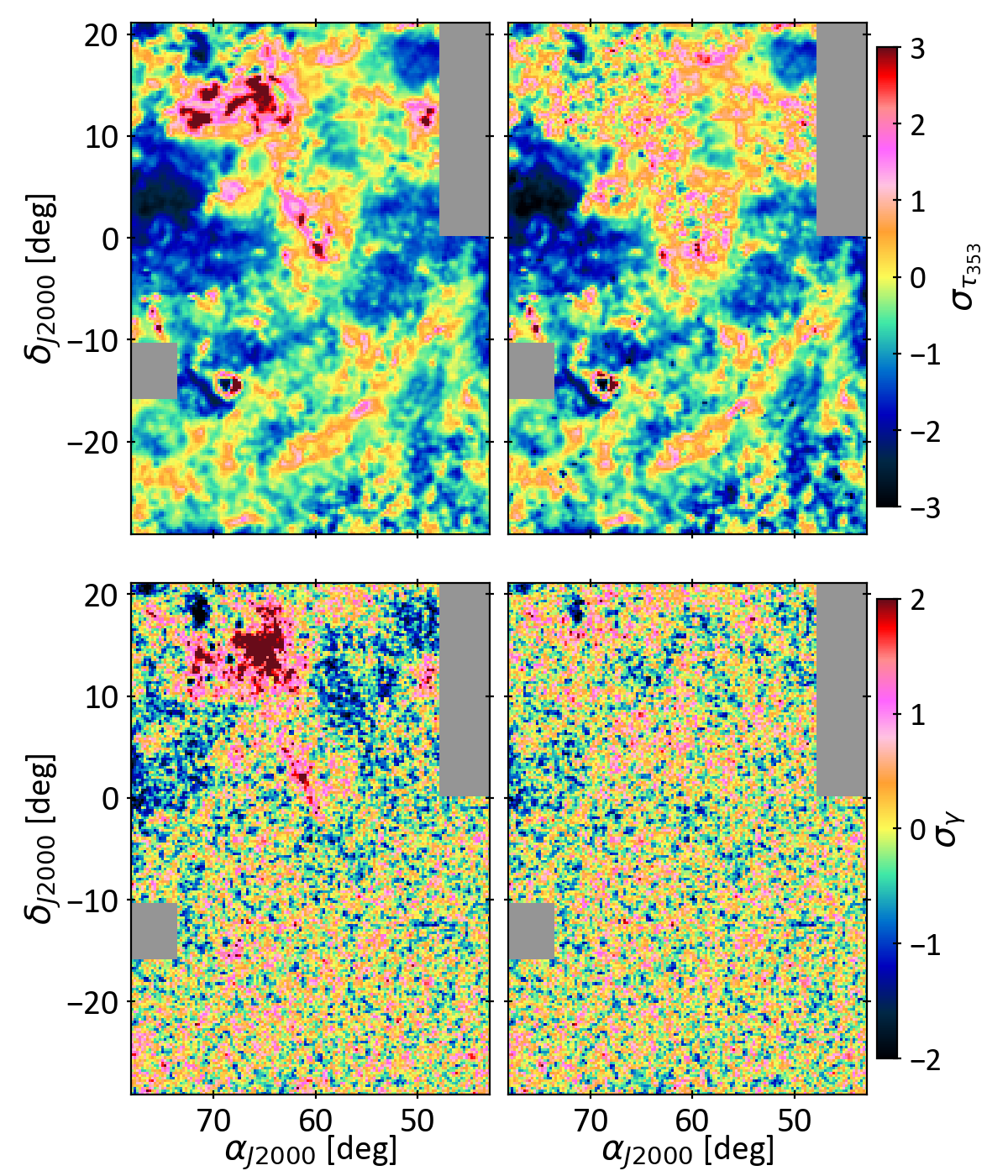}}
		\caption{Residual maps, expressed as data minus best-fit model, in sigma units for the dust optical depth ($\sigma_{\tau}$, \textit{Top}) and the \g-ray counts in the total energy band ($\sigma_{\gamma}, \textit{Bottom}$). The residuals have been smoothed with a 0\fdg15 Gaussian kernel for display. The left- and right-hand side maps show the residuals for models without and with the DNM clouds, respectively.}
		\label{fig:res} 
	\end{center}
\end{figure}

	\subsection{DNM detection}
	\label{sec:DNMdet}
In order to assess the presence of additional gas that is not traced by \hi and CO line intensities, we first performed the dust and \g-ray fits with only the \hi, CO and \halpha templates as gaseous components. The residuals are displayed on the left-hand side of Fig. \ref{fig:res}. The data significantly exceed the model in several extended regions, coherently in dust and in \g rays though the two fits are independent. A coincident variation in CR density and in dust-to-gas ratio is very unlikely and hadronic CR interactions with dust grains produce undetectable \g-ray emission \citep{grenier05}, so this correlated excess has a gaseous origin. The DNM gas distribution is built by correlating and filtering the excess dust and \g-ray signals in the iteration between the dust and \g-ray fits (see Sect. \ref{sec:Iteration}). Adding this template improves the fits quality (see the right-hand side of Fig. \ref{fig:res}). An improvement with a log-likelihood ratio of 5418 when adding the DNM template to the \g-ray fit implies a very significant DNM detection.

The left panel of Fig. \ref{fig:res} shows that most of the DNM gas is within the \northrim. A small contribution comes from the MBM 20 cloud around ($\alpha$,$\delta$) = (68, -14). Keeping the two clouds in the same DNM map assumes a common dust opacity and \g-ray emissivity for both. We have not split the map in two since the DNM mass in MBM 20 is negligible compared to the \northrim one and it does not bias the CR results about the \northrim.

More details on the DNM column densities and mass will be presented in a forthcoming paper, together with a dedicated analysis of MBM 20 at a higher spatial resolution and with an independent DNM template. These results yield much improved dust residuals towards MBM 20 and they give support to keeping the DNM map as a whole for the superbubble analysis. We do not present any CR result for MBM 20 here as its small size requires the higher angular resolution analysis to separate the \g-ray contributions from the different gas phases.

    \subsection{Detection of ionised gas}
    \label{sec:HIIdet}
The \halpha template was added to the models to take into account hadronic CR interactions in the arcs of warm ionised gas. In \g rays, this component is weakly detected at a 4.6~$\sigma$ confidence level over the whole energy band, which is low compared to the ${\sim}74~\sigma$ detection for the DNM. The \halpha template is, moreover, only detected below 2 GeV because of too low photon yields at higher energies from this faint component. Figure \ref{fig:GamModel} shows that this component contributes less than 2 photons per pixel (0\fdg25$\times$0\fdg25) over the entire 0.25–63 GeV energy band. This contribution can be converted into an \nh column density assuming the same \g-ray emissivity per gas nucleon as in the \northrim. The result gives values around 7$\times 10^{19}$~\persqcm that are realistic, but at the limit of detection with the LAT. These values are much lower than the 2 to $30 \times 10^{20}$~\persqcm column densities that characterise the other atomic clouds in the field. We have kept the \halpha template in the models to avoid biasing the other components, but we do not further discuss the physical implications of this faint component.

	\subsection{Best \g-ray and dust fits and jackknife tests}
	\label{sec:jack}
The best-fit model to the \g-ray data includes all the components described in equation \ref{eq:gam} as they have all been significantly detected. The faint \nhi column densities from the Galactic-disc background have not been detected in the dust fit at low spin temperatures because of the overlap with the much more massive \northrim, but all the other components give significant contributions to the best-fit model. 

The $\chi^2$ minimisation and binned maximum-likelihood methods used respectively for the dust and \g-ray fits give access to statistical errors on the best-fit coefficients. They are inferred from the information matrix of the fit \citep{Strong85} and they include the effect of correlations between parameters. Because of the large spatial extents of the cloud complexes and because the gas, dust, and \g-ray distributions are tightly correlated, the statistical errors are small : typically 1–6\% and 1–7\% for the gas parameters of the local clouds in dust and \g rays, respectively. Only the contributions of the Galactic background and \hii template to the \g-ray data have larger uncertainties of 9\% and 17\%, respectively.

We also checked the magnitude of the systematic uncertainties in the linear approximations of the models, hence of spatial changes in the model parameters and/or in the mean level of \hi and CO self absorption. We did so by repeating the last iteration of the dust and \g-ray fits a thousand times over random subsets of the analysis region, namely after masking out 20\% of the pixels with a sum of 3\fdg75-wide, randomly selected squares. In \g rays, the jackknife tests have been performed only for the total 0.25–63 GeV energy band. We have assumed the same relative deviations, $\sigma_q/q$, for the individual energy bands as for the total one since the jackknife sampling tests potential non-uniformities in the models at larger scales than the angular resolution of the data. 

Appendix \ref{ap:Jack} shows the distributions thus obtained for the best-fit coefficients. 
Most of them show Gaussian-like distributions, thereby indicating that the results presented in appendix \ref{ap:coef} are statistically stable and that the average coefficients that describe our models are not driven by subset regions in each cloud complex. A few distributions exhibit long, non-Gaussian tails when the corresponding clouds subtend small solid angles (e.g., the small CO clouds from the \northrim, or the Sun and Moon emissions). The long tails reflect the indeterminacy of the parameter when a large fraction of the non-zero pixels in the corresponding maps is masked. The standard deviations found in the jackknife distributions amount to 1-9\% in dust and to 1-5\% in \g rays, except for the \hi Galactic background (11\%) and the \hii template (23\%) in \g rays.

We quadratically added the 1$\sigma$ fitting errors and the corresponding standard deviations of the jackknife distributions to give the final statistical errors on a given parameter. The results on the best-fit values and final errors are given in appendix \ref{ap:coef}.

	\subsection{Residual maps}
	\label{sec:residual}
The residual maps obtained from the best dust and \g-ray fits for the final iteration are presented in Figure \ref{fig:res} (\textit{right-hand side}). 
When including all gaseous components, in particular the DNM one, we see that our linear model provides an excellent fit to the \g-ray data (\textit{bottom right}). The residuals are consistent with noise except, marginally, towards a compact CO cloud near ($\alpha,\delta$)=(72,18) where the model slightly over-predicts the data. This may be due to a lower \nhd-to-\wco conversion factor in this cloud compared to the larger values that characterise the other, more tenuous, clouds in the \northrim molecular map. Such variations have been observed in other molecular clouds \citep{Remy17}.  

The dust residual map (\textit{top right}) presents structured deviations. The ring-like excess and central deficit towards MBM 20 is due to the 0\fdg25 sampling resolution of the present analysis that is not sufficient to capture the multi-phasic structure of this very compact molecular cloud. These residuals vanish in the higher angular resolution analysis that will be presented in the forthcoming paper. The other excesses, towards the \northrim and \westrim, are likely due to the non-linear increase in dust opacity in the dense parts of these clouds and will be discussed in the forthcoming paper. 

 \begin{figure}
	\begin{center}
		\resizebox{\hsize}{!}{\includegraphics{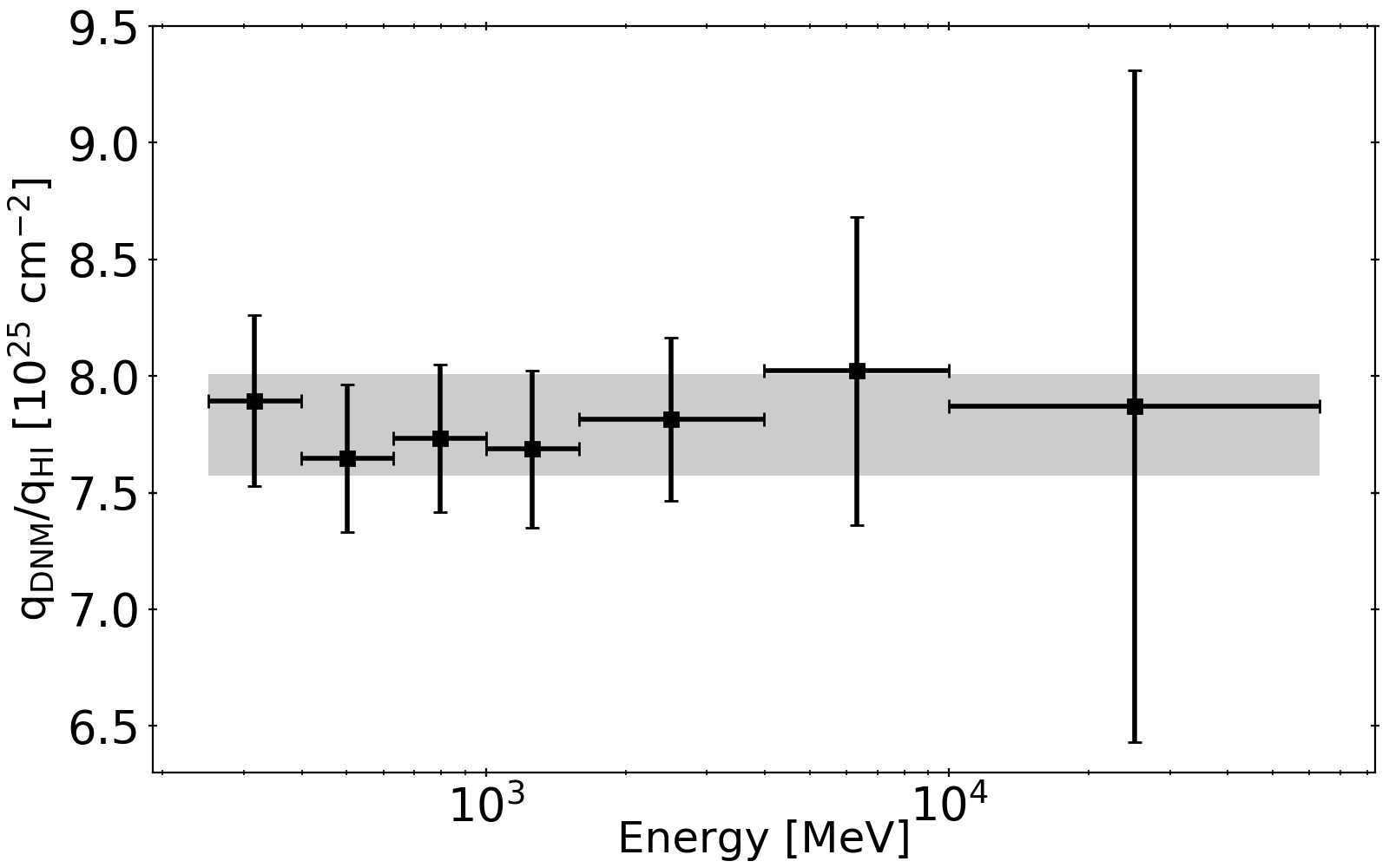}}
		\caption{Ratio of the DNM to \hi emissivity factors in the \northrim as a function of energy, for a spin temperature of 100~K. The grey band gives the $\pm 1\sigma$ confidence range on the ratio obtained for the full energy band. The constant ratio indicates that the same CR spectrum permeates the \hi envelope and DNM phase.}
		\label{fig:qDNM} 
	\end{center}
\end{figure}
 \begin{figure}
	\begin{center}
		\resizebox{\hsize}{!}{\includegraphics{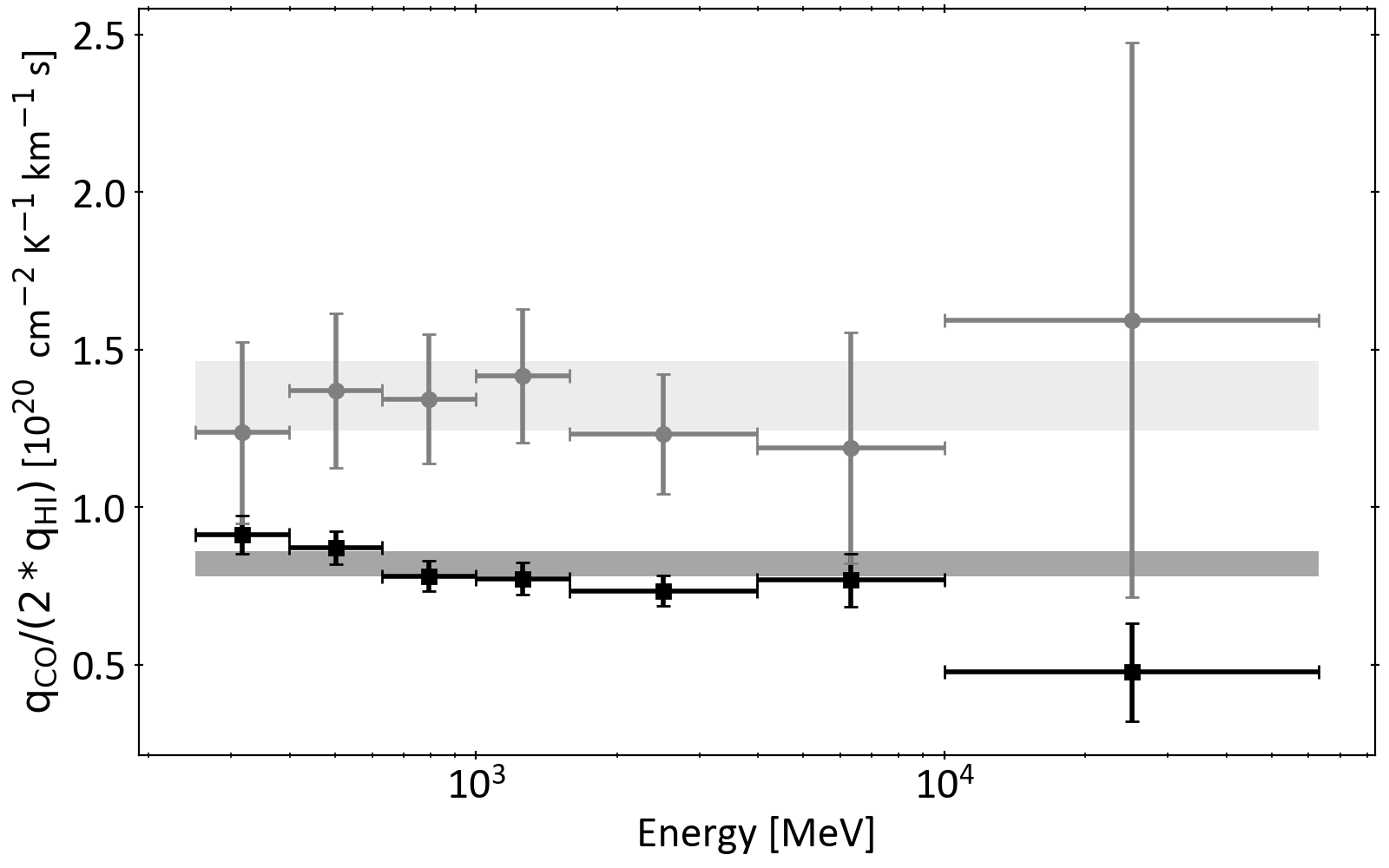}}
		\caption{Ratio of the CO to \hi emissivity factors as a function of energy, for a spin temperature of 100~K, for two sets of molecular clouds, in the \northrim of the  superbubble (dark) and in the North-Taurus-Cetus clouds (light). The bands give the $\pm 1\sigma$ confidence range on the ratios obtained for the full energy band. The constant ratios indicate that the same CR spectra permeate the atomic and molecular phases of the clouds.}
		\label{fig:qCO} 
	\end{center}
\end{figure}

	\subsection{Cosmic-ray penetration in the different gas phases}
  
For a given cloud, spectral changes in the ratio between the \g-ray emissivities measured in the DNM and CO-bright gas and that in the diffuse atomic envelope can reveal changes in the underlying CR distribution as the particles penetrate into the molecular phases. One expects low-energy CR (below 10 GeV) to be depleted as they diffuse away on self-excited MHD turbulence in the cloud envelope. The turbulence is triggered by the CR density gradients induced by severe hadronic losses in the cloud interior \citep{Skilling76,Ivlev2018}.  

Figure \ref{fig:qDNM} shows that we found a stable DNM to \hi emissivity ratio as a function of energy in the massive \northrim of the superbubble. It implies that the same CR spectrum permeates the atomic and DNM phases even though the gas is denser and less ionised at the \hi-\hd transition than in the warm \hi envelope. The \northrim DNM mass is large (of order 1.2$\times 10^4$~M$_{\odot}$ at 160~pc), but the DNM gas column densities of 2 to 20$\times 10^{20}$~\persqcm are one to two orders of magnitude below the critical values for significant low-energy CR depletion \citep{Ivlev2018}. 

Likewise, for the \northrim and North-Taurus-Cetus clouds, we found no spectral evolution of their respective CO to \hi emissivity ratios. The weak down-going trend seen for the \northrim ratio is not statistically significant.
The data is consistent with the same CR spectrum flowing through the \hi envelope and CO-bright gas, the latter having charateristic volume densities of a few $10^3$~\percc and magnetic field strengths of 1-10~$\mu$G \citep{Crutcher12}. The molecular gas column densities in these clouds are indeed far below the critical values for low-energy CR depletion. These results confirm the same trends found in other nearby clouds \citep[][and references therein]{Grenier15,Remy17}. 

		\subsection{Cosmic rays in the Orion-Eridanus superbubble}
		\label{sec:qSB}
\begin{figure}
	\begin{center}
    	\resizebox{\hsize}{!}{\includegraphics{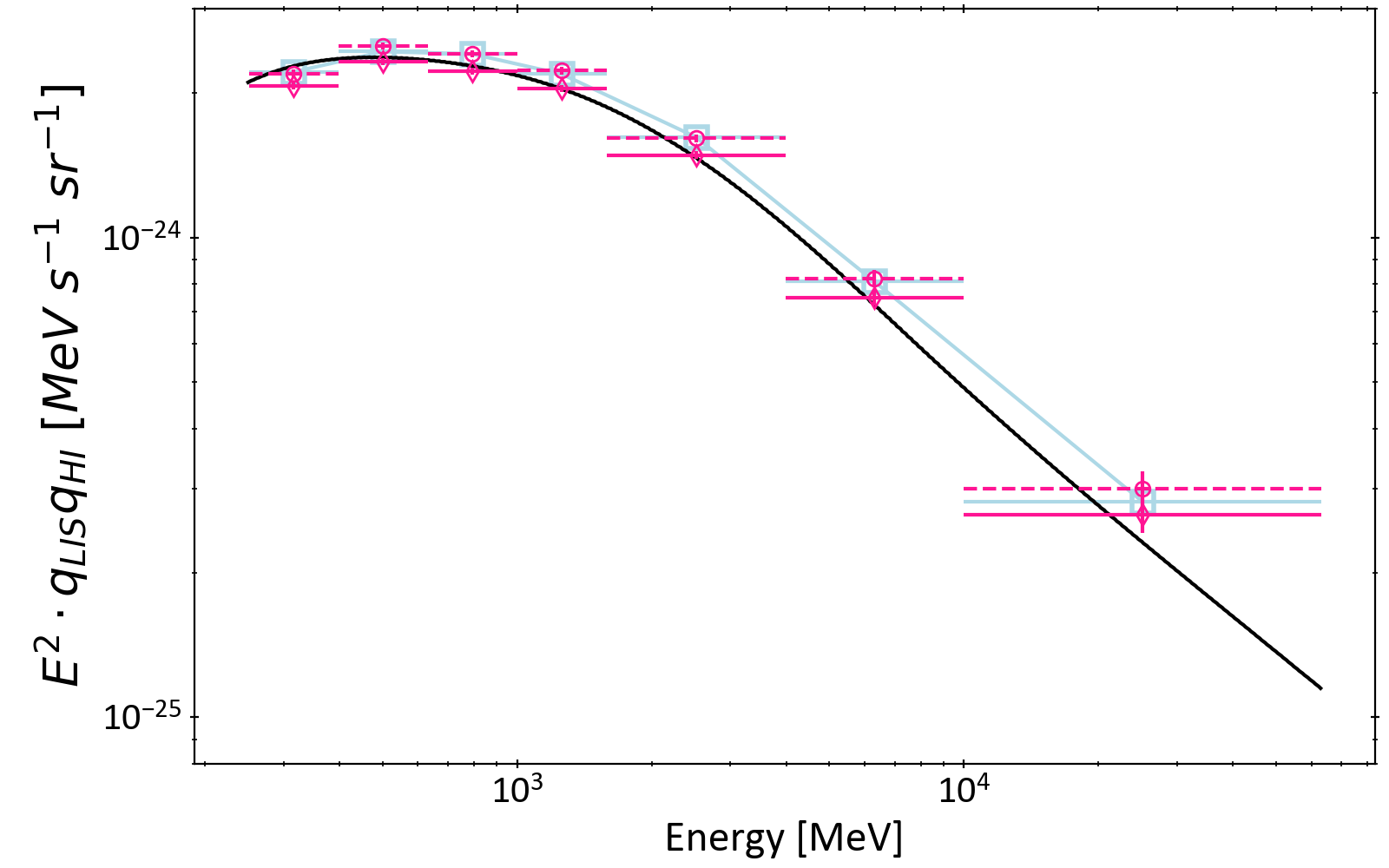}}
    	\caption{Average spectral energy distribution of the \g-ray emissivity per gas nucleon, \qsb, found in the atomic gas of the superbubble for an \hi spin temperature of 150~K (magenta diamonds, solid lines) and for the optically thin \hi case (magenta circles, dashed lines). The blue squares show the average emissivity, \qlis, found in the local ISM for a spin temperature  of 140~K \citep{Casandjian15}. The black line gives the \g-ray emissivity per gas nucleon inferred from the CR spectrum measured in the Solar System (see text).}
    	\label{fig:qsb}
	\end{center}
\end{figure}

We first probed the average \g-ray emissivity, \qsb, of the gas in the superbubble by gathering the four superbubble clouds into a single component. This emissivity was derived for a spin temperature of 150~K in order to compare with the average emissivity, \qlis, derived in the local ISM for a spin temperature of 140~K \citep{Casandjian15}. The result is displayed in Fig. \ref{fig:qsb}. 
The superbubble emissivity presents the same spectral energy distribution (SED) as the local ISM average. The 6\% difference is highly significant (12$\sigma$) over the entire energy band. Figure \ref{fig:qsb} shows, however, that the difference is consistent with the systematic uncertainties in the \hi optical depth of the different clouds that have served to perform both measurements. The difference is also compatible with the $\pm$9\% dispersion found in other nearby clouds for the same reason \citep{Remy17}. The amplitude and spectral shape of the mean \qsb emissivity in the superbubble implies that the CR spectrum pervading this region is not modified, in particular at high energy, despite the perturbed conditions and past supernova history of the superbubble.

We compared the \g-ray emissivity spectrum in the superbbuble with that inferred from the CR spectrum measured in the Solar System. To do so, we used the parametrisations of the CR proton and helium spectra proposed by \citet{Corti16} on the basis of the \textit{Voyager 1}, AMS-02, \textit{PAMELA}, and  \textit{BESS} data and the use of an updated, rigidity-dependent, force-field approximation to account for the solar modulation. We also used the CR electron spectrum inferred from the \textit{Voyager 1}, AMS-02, \textit{PAMELA}, and \textit{Ulysses} data and the HELMOD code for CR propagation inside the heliosphere \citep{Boschini18}. We used the hadronic-interaction cross sections from \citet{Kamae2006} and the multiplication factors to take elements heavier than H into account from \citet[][see also \citeauthor{Kachelriess14} \citeyear{Kachelriess14}]{Mori2009}. The result for the \g-ray emissivity per gas nucleon in hadronic and bremsstrahlung interactions is presented in Fig. \ref{fig:qsb}. It shows that the CR spectrum diffusing through the superbubble is compatible within the uncertainties with the particle distribution passing through the Solar System. 

\begin{figure*}
	\centering
	\includegraphics[width=17cm]{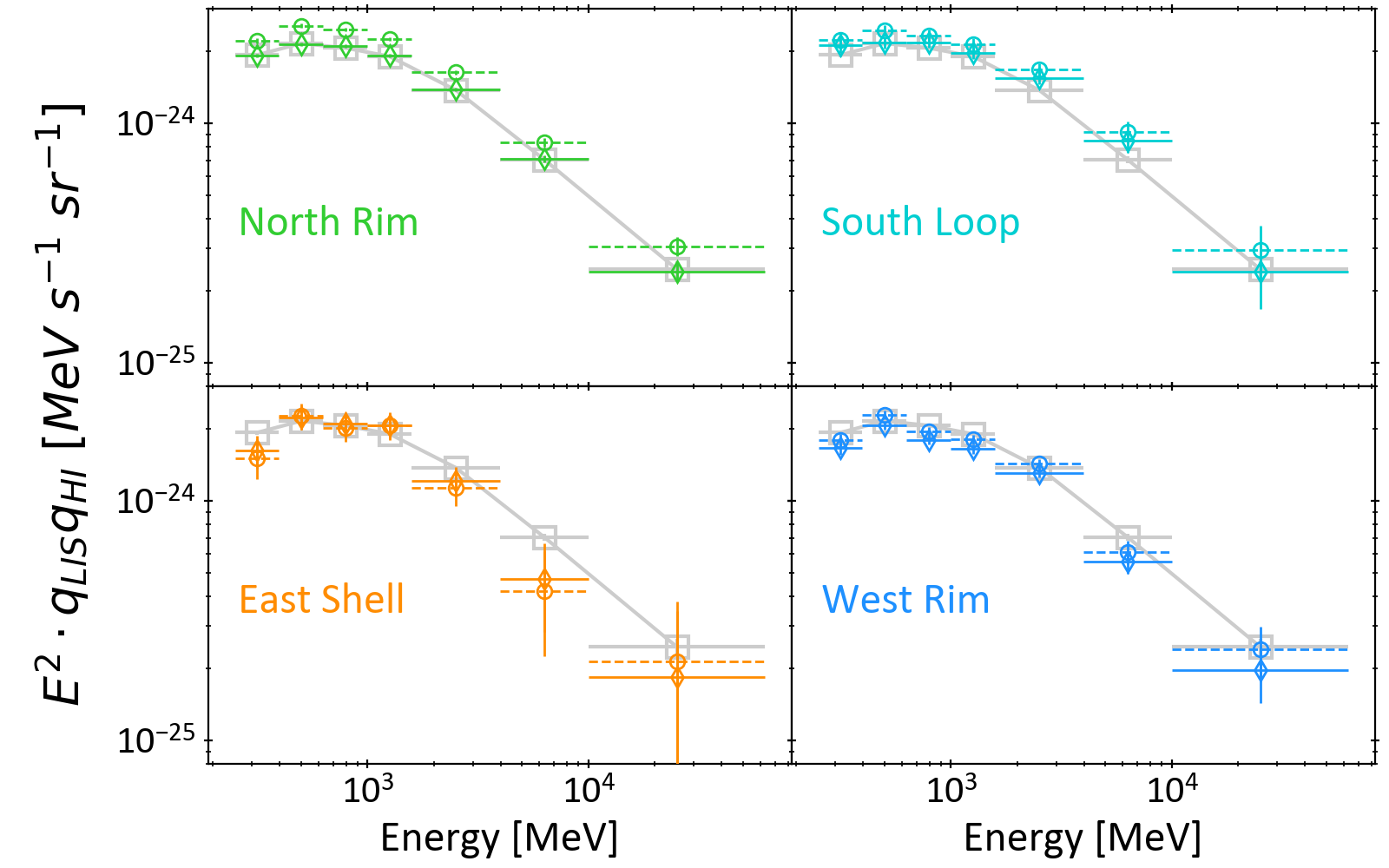}
	\caption{Spectral energy distributions of the \g-ray emissivity per gas nucleon found in different atomic clouds  associated with the superbubble, for an \hi spin temperature of 100~K favoured by the fit (solid, diamonds) and for the optically thin \hi case (dashed, circles). The grey squares show the average emissivity in the superbubble, \qsb, for an \hi spin temperature of 100~K.}
	\label{fig:qMos}
\end{figure*}

We probed the uniformity of the \g-ray emissivity across the superbubble. The emissivity SED found in individual \hi clouds associated with the superbubble are shown in Figure \ref{fig:qMos} for the spin temperature of 100~K favoured by the \g-ray fit for this region. They are compared with the average \qsb emissivity in the superbubble in order to search for variations. The emissivity SED in the \northrim and \southloop are consistent with \qsb at all energies within the uncertainties. The slightly lower emissivity found in the \eastshell at the lowest and highest energies is not statistically significant (0.6$\sigma$). The 8\% lower emissivity in the \westrim differs from the \qsb average at a 5.6$\sigma$ confidence level, but the result for the optically thin case shows that the small difference may stem from cloud to cloud variations in the \hi optical depth correction. 

		\subsection{Cosmic rays in the Eridu cirrus}
        \label{sec:qEridu}
\begin{figure}
	\begin{center}
		\resizebox{\hsize}{!}{\includegraphics{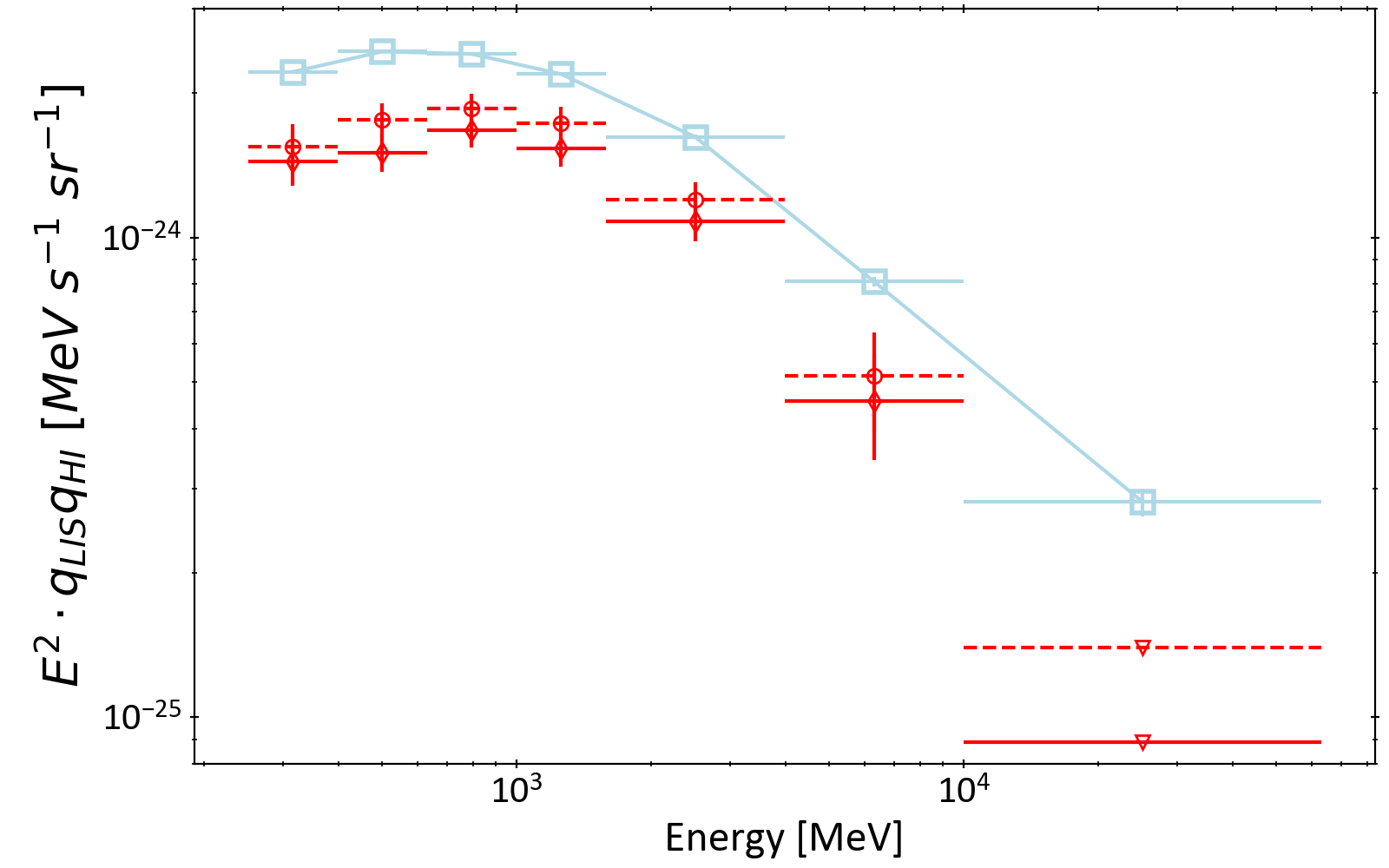}}
		\caption{Spectral energy distribution of the \g-ray emissivity per gas nucleon found in the atomic gas of the Eridu cirrus, for an \hi spin temperature of 100~K favoured by the fit (solid, diamonds) and for the optically thin \hi case (dashed, circles). The blue squares show the average emissivity, \qlis, found in the local ISM for a spin temperature  of 140~K \citep{Casandjian15}.}
		\label{fig:qCirrus}
	\end{center}
\end{figure}

The Eridu cirrus cloud lies outside the superbubble, at an altitude of about 200-250~pc below the Galactic plane. Its distance was derived in \citetalias{Joubaud19} using the dust reddening map of \citet{Green18}. The SED of the \g-ray emissivity we found in this cloud is displayed in Fig. \ref{fig:qCirrus}. Compared with the local ISM \qlis average, we found a highly significant (14$\sigma$) drop in emissivity by 34\%. This difference cannot be attributed to \hi optical depth corrections as the optically thin case, which sets the minimum amount of gas, gives an upper limit to the emissivity that is still 25\% below \qlis. We do not detect any significant amount of DNM gas in the Eridu direction, but adding any missing gas would imply an even smaller \g-ray emissivity per H nucleon to produce the same \g-ray intensity. Confusion in the cloud separation in velocity is not possible as it is the only \hi structure seen towards these directions. At Galactic latitudes near -50\degr, it is indeed too far from the Galactic plane for significant background gas contamination. 

Given the modest gas column densities in the cloud ($1-5 \times10^{20}$~cm$^{-2}$), we checked the influence of large-scale \g-ray structures from the IC and isotropic emission components. Varying those by $3 \sigma$ around their best-fit spectra yields a maximum variation of 6\% for the Eridu emissivity spectrum. We also assessed the influence of the spatial distribution of the Galactic IC emission across the region by trying different GALPROP models. The radial distribution of CR sources in the Galaxy as well as the halo size directly influence the IC intensity gradient across the analysis region, but in a limited way. Relative variations in intensity do not exceed 10\% if we compare the case of a steep Galactic CR gradient (with CR sources following the peaked distribution of OB stars and for a small halo height of 4 kpc) and the case of a shallower CR gradient (with CR sources following the distribution of supernova remnants and for a halo height of 10 kpc) \citep{Ackermann12}. The Eridu emissivity spectrum changes by less than 3\% if we replace the present IC model by one of those more extreme cases. These tests show that the spatial distribution of the gas in the Eridu cloud is significantly detected in the \g-ray fit, with little influence from the large-scale structures in non-gaseous components.

\section{Discussion}
\label{sec:Discussion}

\subsection{\hi optical depth correction}
Among the set of spin temperatures we used to account for \hi optical depth corrections, the maximum log-likelihood values favour a temperature of 100~K, as shown in appendix \ref{ap:Tspin}. This temperature is lower than in other nearby clouds \citep{PlanckXXVIII15,Remy17}. It is also low compared to the values closer to 300~K expected for \nhi column densities in the range of $(2-30) \times 10^{20}$~\persqcm sampled here. Such column densities only partially shield the gas from the interstellar UV and soft X-ray radiation field ($0.1 \lesssim A_V \lesssim 1.5$) \citep{Kanekar11,Bohlin78}.
Our model assumes a single spin temperature over the entire line of sight line (monophasic \tspin) and in the different clouds. The latter implies that the spin temperature determination is biased by the most extended clouds because of their preponderant weight in the fit. Among the different clouds in the field, the main \hi contribution comes from the \northrim which, being rich in DNM gas at the dense \hi and diffuse \hd transition, likely contains a large fraction of cold neutral medium structures. The latter have spin temperatures well below 200~K \citep{Murray15}.  

It should, however, be noted that, due to the relatively low column densities involved here, the optical depth correction has only a small impact of order 10\% on the CR spectrum derivation. The relative differences in \g-ray emissivities between the \tspin=100~K and the optically thin cases are 1, 10, 10, 17, and 14\% for the \eastshell, \westrim, \southloop, \northrim, and Eridu cirrus, respectively.

\subsection{The cosmic-ray flux in the superbubble}

The CR spectrum, probed here at energies between approximately 2 and 500 GeV, appears to be uniform across the solar neighbourhood and the older end of the Orion-Eridanus superbubble. It is consistent with the direct CR measurements in the solar system. This energy spectrum in $E^{-2.7}$ indicates that CR propagation near the superbubble environment should be dominated by the same energy-dependent mode of diffusion as in the Galaxy at large. We note that the present \g-ray emissivity measurements mainly probe the CR flux in the outer shells of the superbubble and not directly in its hot interior plasma. The latter yields too low column densities to be detectable in \g rays with the LAT, and even the denser warm ionised gas is not firmly detected in our analysis. CR acceleration or re-acceleration may then take place in the superbubble interior or in the more active zone at the younger end of the superbubble (not studied here), but it leaves no detectable trace in the outer shells closer to us that we have studied. The emissivity measurement in the \eastshell suggests that the CR flux inside the superbubble is consistent with that of the local ISM as this cloud is considered to be inside the hot gas.

\begin{figure}
	\begin{center}
		\resizebox{\hsize}{!}{\includegraphics{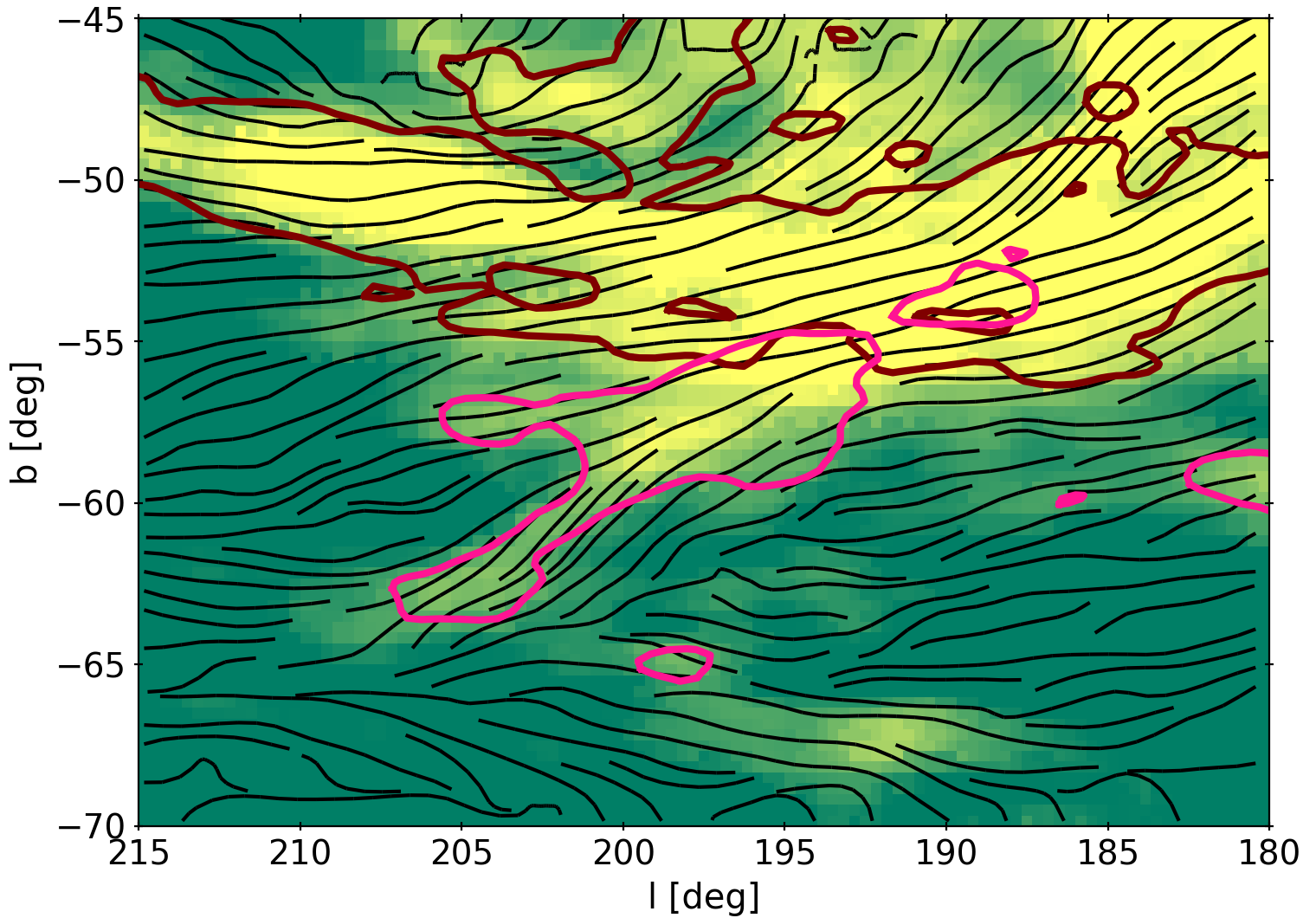}}
		\caption{\nhi column density map in Galactic coordinates. The brown contour delineates the \westrim and the \southloop of the superbubble at 2.6 $\times 10^{20}$ cm$^{-2}$ and the magenta countour delineates the Eridu cirrus at 3.1$\times 10^{20}$ cm$^{-2}$. The black lines give the orientation of the plane-of-the-sky magnetic field inferred from the dust polarisation observations at 353 GHz with \Planck.}
		\label{fig:bfield}
	\end{center}
\end{figure}

Galactic CR may not penetrate deep into the superbubble if there is strong turbulence inside it \citep{Tolksdorf19}. Conversely,
for an inefficient accelerator as the Orion-Eridanus superbubble seems to be, CR can pass through the hot plasma without significant changes. Yet, the magnetic field lines appear to be wrapped around the superbubble, having been ordered and compressed along the outer rim by the shock expansion. Figure \ref{fig:bfield} shows the orientation of the plane-of-the-sky magnetic field (\bsky) in the \westrim, where strengths up to 15~\muG have been inferred \citepalias{Joubaud19}. CR entering the shell will mainly diffuse along these field lines.

Transport perpendicular to the average field lines is possible with the field line random walk \citep{Jokipii66}. Particles can also drift across the field because of large-scale variations in magnetic direction or strength, or by scattering off small-scale MHD structures \citep{Desiati14}. Perpendicular transport is, however, always much slower than parallel transport. Even if advected with Alfvén waves, the bulk of CR would stream at the Alvén speed along the large-scale field lines. The ordered magnetic field in the outer shell could then limit the mixing between the internal and external CR populations. On the other hand, we expect the older part of the superbubble to be filled with numerous weak secondary shocks, reflected off the corrugated outer shell and by interior cloudlets, as observed in simulations \citep{Kim17}. This turbulence, scaling as $k^{-2}$, may help CR to diffuse in an energy-independent way across the outer shell despite its large-scale magnetic configuration \citep{Bykov87}. 

In this context, the comparison between the \g-ray emissivities of the \eastshell and \westrim can provide insight into CR transport. 
The \eastshell is considered to be inside the hot plasma phase because of the lack of X-ray absorption against the bright interior plasma \citepalias{Joubaud19}, but the distance uncertainties may as well place it as an outer-rim shell in the lower, rear part of the superbubble. Future absorption studies, in X rays and with dust, will be crucial to precisely locate this cloud and its \g-ray flux measurement with respect to the edge of the superbubble. 

The lack of a CR enhancement or hardening in the Orion-Eridanus superbubble is in marked contrast with the case of the Cygnus X cocoon where a bright $E_{\gamma}^{-2.1}$ differential emissivity spectrum has been detected from 1 to 100 GeV \citep{Ackermann11}.
We discuss below possible origins of the apparent inefficiency of the Orion-Eridanus superbubble as a potential CR (re)-accelerator. 

    \subsubsection{Intermittent shock acceleration}
    \label{sec:accel}
The first possible origin concerns the acceleration conditions inside the superbubble. The acceleration efficiency by intermittent shocks and magnetic turbulence depends on the spectral energy density of the magnetic fluctuations that results from the complex interplay between energy cascading, wave dissipation, and CR-driven instabilities. The outcome is not known and the spectrum is often modelled as a $k^{-b_T}$ power law in wavenumber, scaling from the injection scale $L^*$. For strong magnetic turbulence, with an amplitude $\eta_T {=} \langle\delta B^2\rangle/(B^2{+}\langle\delta B^2\rangle) \approx 1$ close to saturation, the particles with gyroradius $R_g$ and velocity $v$ diffuse isotropically with a diffusion coefficient $D = D_B \,(R_g/L^*)^{1-b_T} \eta_T^{-1}$ that largely exceeds the Bohm value $D_B = R_g v/3$. The stellar density in the superbubble and the results of superbubble MHD simulations \citep{Ntormousi17} suggest a typical injection scale $L^* \approx 10$~pc and interior magnetic-field strengths of order 1--3 \muG. In these conditions, the diffusion coefficient spans from 
\begin{equation}
D= 3.2 \times 10^{26}~\mathrm{cm}^2~\mathrm{s}^{-1}~\left(\frac{E}{10\,{\rm GeV}}\right)^{1/2} \left(\frac{L^*}{10\,{\rm pc}}\right)^{1/2} \left(\frac{B}{1\mu {\rm G}}\right)^{-1/2} \eta_T^{-1}
\end{equation}
for a Kraichnan-like spectrum $b_T=3/2$ \citep{Perez12}, to a tenfold larger value
\begin{equation}
D'= 3.1 \times 10^{27}~\mathrm{cm}^2~\mathrm{s}^{-1}~\left(\frac{E}{10\,{\rm GeV}}\right)^{1/3} \left(\frac{L^*}{10\,{\rm pc}}\right)^{2/3} \left(\frac{B}{1\mu {\rm G}}\right)^{-1/3} \eta_T^{-1}
\end{equation}
for a Kolmogorov-type index $b_T=5/3$. 

These values are respectively 200 and 20 times smaller at 10~GeV than the average uniform coefficient inferred in the Milky Way using the GALPROP propagation code and elemental spectra from AMS-02 and \textit{ACE/CRIS} \citep{Gully19}. The size of the TeV haloes found around the nearby Geminga and PSR B0656+14 pulsars requires slow diffusion between 1.5 and $7.0 \times 10^{26} \,(E/10\,{\rm GeV})$~cm$^2$/s \citep{Abeysekara17}. These values compare well with our superbubble estimates and they apply to the same distance range and anticentre direction. The TeV data is best described with a 3~\muG ambient field and a coherence length $\lesssim 5$~pc of the magnetic turbulence \citep{LopezCoto18}. These conditions are close to the values used for the superbubble. Yet, even though Geminga is born in the superbubble \citep{Pellizza05}, it now lies 100~pc outside it, as do PSR B0656+14, and the extent of the slow diffusion zone around the pulsars is still debated. It could be restricted to the relic pulsar wind nebula \citep{Tang19} or to a region ${\lesssim} 100$~pc controlled by streaming instabilities \citep[][ and references therein]{Gully19}. So, one cannot directly compare diffusion lengths inside the superbubble and around the two pulsars.

In the Kraichnan case, the confinement time, $L^2/[6D]$, in the $L{\approx}200$-pc-wide bubble is long enough ($>$ 3.9 Myr) to accelerate particles from supra-thermal energies up to a moderate maximum energy of 30-80 GeV set by $D(E_{max}) \approx u_{sh} L^*/3$ \citep{Bykov01} if we assume typical velocities $u_{sh}\approx 500$~km/s for the shocks and large-scale turbulent flows in the old part of the superbubble. The case of Kolmogorov turbulence  is much less favourable. The maximum energy is limited to a few GeV as the particles diffuse out of the system in less than one Myr.  \citet{Ferrand10} reached the same conclusion about a fast escape restraining CR acceleration in a 300~pc-long superbubble filled with 1~\muG plasma around Orion OB1. 

Acceleration is slow in the weakly magnetised interiors of superbubbles, but the spectral index of the turbulence at the small scales that govern how fast the particles diffuse away also plays an important role in the overall acceleration efficiency. Hence the Orion-Eridanus superbubble may be less efficient than the Cygnus X one because of a weaker magnetic field inside the bubble and/or a steeper turbulence spectrum. MHD superbubble simulations can inform us on the turbulence spectrum inside CR-inefficient bubbles like Orion-Eridanus, but one needs to include CR feedback processes to model the magnetic turbulence inside active systems like Cygnus X \citep{Bykov14}. The feedback involves streaming instabilities, resonant wave-CR interactions, and the energy transfer from the plasma to CR.

    \subsubsection{Time evolution}
    \label{sec:time}
The non-linear models by \citet{Bykov01} and \citet{Bykov14} include intermittent shocks, a description of the correlations between the shocks and the long-wavelength turbulent motions of the compressible plasma, and energy conservation between the background plasma and CR population. The results show that the Orion-Eridanus system has the potential to accelerate CR up to a few hundred GeV \citep[see Fig. 2 of ][]{Bykov01}. The CR spectrum, however, evolves considerably with time, so one should consider the late development stage of the Orion-Eridanus superbubble. Even though the supernova rate is roughly constant over tens of Myr for stellar populations following the local initial mass function \citep{Ferrand10}, the models predict a soft-hard-soft evolution of the output CR spectrum with time. The late softening is due to the damping of the shocks and of the large-scale compressible turbulence because of the energy transfer to CR. It is not due to adiabatic losses to the superbubble expansion as the models are quasi-static because of the slow expansion. This is a valid assumption given the present expansion velocity of 20~\kmpers of the Orion-Eridanus superbubble \citepalias{Joubaud19}.
Radiative losses are also unimportant in the rarefied plasma inside the superbubble \citepalias[$n_H = 0.005 - 0.05$~cm$^{-3}$,][]{Joubaud19}. 
Given its age, the Orion-Eridanus system is now in the late softening stage tending towards $p^2 f(p) \propto p^{-3}$. The hardest period (${\propto} p^{-2}$) occurred earlier, after 1~Myr of evolution in the non-linear model of \citet{Bykov01} and after 10~Myr in the linear version of \citet{Ferrand10}.

One can further test this spectral ageing by investigating the younger part of the superbubble, closer to Orion. 
The stellar history in the superbubble, as pictured by the stream of blue stars spanning from the closest wall ($\sim$20-Myr old stars) down to Orion ($\sim$1-Myr old stars), indicates that the energy injection progressed in space over a period of several Myr \citep{Bouy15}. The X-ray emission from the hot interior plasma, tracing its current energy content \citep{krause14}, presents a clear hardening towards Orion \citep[][\citetalias{Joubaud19}]{Ochsendorf15}. The energetic sources have shifted to larger distances with time. 
But the gradual propagation of massive-star formation may not readily explain the CR contrast between Orion-Eridanus and Cygnus X since star formation has also progressed across Cygnus X over a comparable period, between ${\gtrsim}10$ and 1~Myr ago \citep{Berlanas18}. 

\subsubsection{Stellar content}
A notable difference between the inefficient Eridanus and efficient Cygnus X accelerators is the presence of a super-massive cluster in the latter : Cygnus OB2 still gathers 143 supernova progenitors
\citep[$>8~M_{\odot}$,][]{Wright2015}. A similar total of 62+48 is found across Orion OB1 and the blue stream \citep{Bouy15}, but Cygnus OB2 has produced ${\sim} 6$~SN/Myr for the last Myr \citep{Lingenfelter18}, whereas 1~SN/Myr has exploded on average in Orion-Eridanus over the last 12~Myr \citep{Voss10,Bouy15}.
The higher space density and frequency of supernovae in the last Myr in Cygnus X may have favoured CR acceleration compared to Orion-Eridanus. Colliding shocks, in particular, are more frequent in a rich and compact cluster such as Cygnus OB2. The interaction of a supernova shock with one or several fast approaching and massive stellar winds can rapidly accelerate CR to PeV energies with a hard $p^2 f(p) \,{\propto} p^{-1}$ spectrum \citep{Bykov13}. Their diffusion to $>30$~pc distances in less than $10^4$ years after they escape from the colliding shocks would produce a hard CR cocoon in Cygnus X that has not recently occurred in Orion-Eridanus where the stellar density is lower.
Future Gaia studies will help to better constrain the stellar history of Orion-Eridanus as, unfortunately, many of the Orion OB stars are too bright to be part of the present Gaia DR2 survey \citep{Gaia2018}.

\subsubsection{Superbubble size}
As mentioned in the introduction, the massive OB association G25.18+0.26 may power a cocoon of hard CR \citep{Katsuta17}. At a distance of 7.7~kpc, its projected size of $210\times 170$~pc compares with that of Orion-Eridanus, so the different CR behaviours of the two systems do not simply relate to their comparable sizes. The models indeed predict that the efficiency of the power transfer from turbulence to CR always exceeds 10\% after the first Myr, irrespective of the superbubble size (from 50 to 220~pc). The larger systems are more efficient at late times because of a reduced escape rate \citep{Bykov01,Bykov14}, but the difference is small (factor of ${\sim}2$). 
The next generation of \g-ray telescopes, such as CTA, will play a key role in extending the present sample of superbubble observations and in assessing the influence of their size and of their stellar-cluster content on CR acceleration. With higher resolution imaging, the investigations for hard CR should be directed towards sources like Westerlund~2, barely resolved by H.E.S.S. \citep{HESS11Wd2}, towards the confused case of Westerlund~1 where the GeV and TeV emissions appear to have different origins \citep{Ohm13}, towards NGC 3603 which was not resolved by Fermi LAT in the 4FGL catalogue\footnote{https://fermi.gsfc.nasa.gov/ssc/data/access/lat/8yr\_catalog/}, or towards the bright Arches, Quintuplet, and Cl$*$1806-20 stellar clusters near the Galactic Centre, as suggested by \citet{Ackermann11,HESS12Cl1806,HESS18,Aharonian2019}.

\begin{figure}
	\begin{center}
		\resizebox{\hsize}{!}{\includegraphics{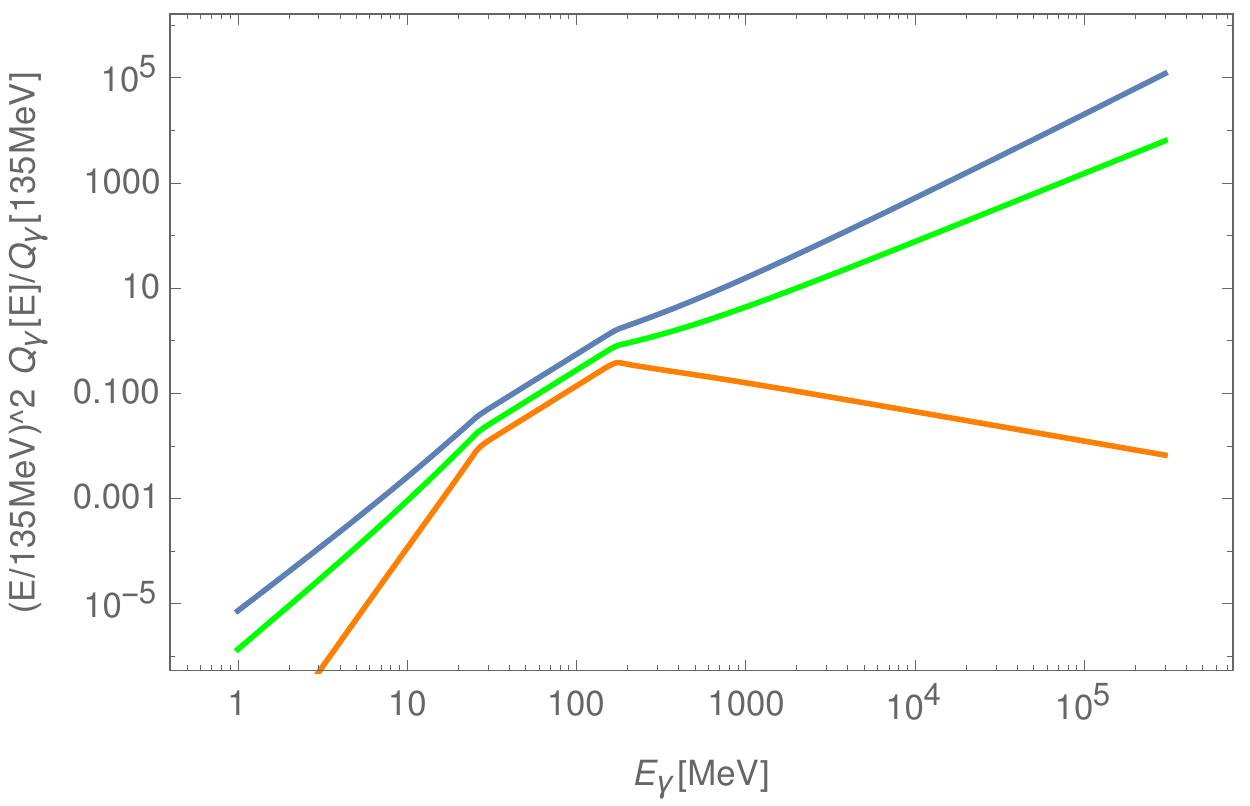}}
		\caption{Theoretical spectral energy distributions of the gamma rays resulting from pion decay inside a turbulent superbubble for different values of the acceleration efficiency parameter $\xi$ \citep[according to][]{Tolksdorf19}. The efficient case ($\xi=0.1$) is in blue, the intermediate one ($\xi=1$) in green and the inefficient case ($\xi=10$) in orange. All distributions are normalised to one at the pion mass. For a better readability, the SED for $\xi=1$ and $\xi=10$ are divided by a factor of two and four, respectively.}
		\label{fig:Tobias}
	\end{center}
\end{figure}

\subsubsection{Cosmic-ray re-acceleration}
We found no \g-ray evidence of re-acceleration of Galactic CR in the  Orion-Eridanus superbubble. The bubble expansion has swept-up, compressed, and wrapped the ambient ISM magnetic field to the point where the field lines appear to be mostly tangent to the outer rim and 
perpendicular to the shock normal \citep[][\citetalias{Joubaud19} and Fig. \ref{fig:bfield}]{Soler18}. This was considered as an unfavourable configuration for CR acceleration by the outer shock wave \citep{Ellison95}. Recent simulations indicate, however, that the magnetic amplification by streaming instability can rearrange highly inclined fields ($\lesssim 70$\degr) to quasi-parallel configurations that allow particle injection into the shock for diffusive re-acceleration \citep{Caprioli18}. The downstream medium may also be rich in supra-thermal particles due to the series of past supernova remnants in the superbubble and they can get re-accelerated.

\citet{Tolksdorf19} have analytically studied stochastic re-acceleration in the turbulent interior of a spherical superbubble, including spatial and momentum diffusion as well as pion losses. The acceleration potential is captured by the ratio of the acceleration time scale, $\tau_A \sim 9D(p)/v_A^2$, over the diffusive escape timescale, $\tau_D \sim R_{SB}^2/D(p)$, with $R_{SB}$ the superbubble radius, $v_A$ the Alfven velocity in the interior plasma, and $D(p)$ the spatial diffusion coefficient estimated in Sect. \ref{sec:accel}. 
The ratio $\xi(p) = [\tau_A/\tau_D]^{1/2} = 3 D(p) / [v_A R_{SB}]$ indicates if CR have spent enough time in the turbulent region to be re-accelerated before diffusing out. Pion losses play a negligible role in the Orion-Eridanus interior as their timescale is orders of magnitude longer than $\tau_A$ and $\tau_D$. 

\begin{figure}
	\begin{center}
		\resizebox{\hsize}{!}{\includegraphics{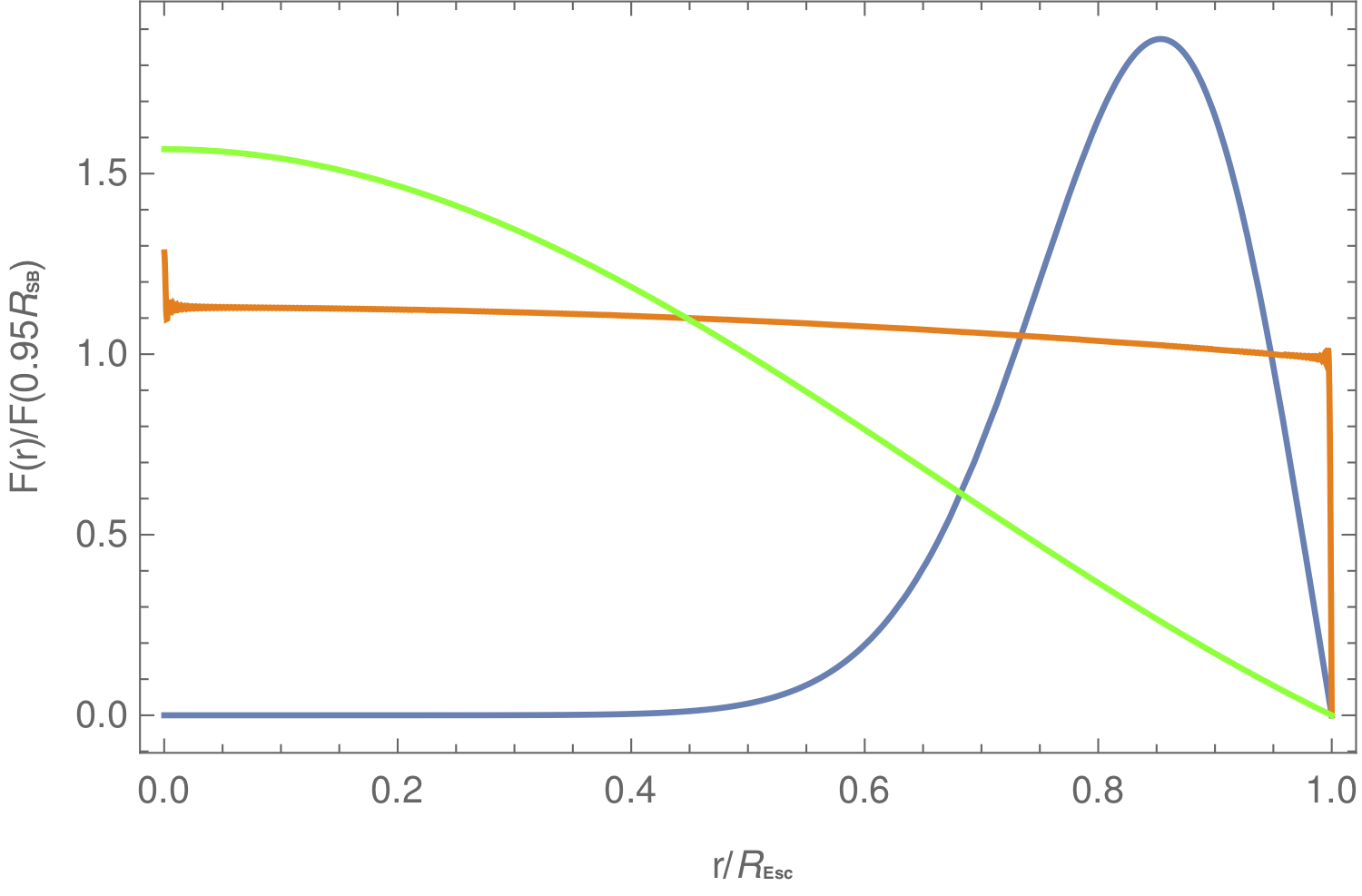}}
		\caption{Theoretical steady-state proton distribution, integrated between 10 GeV and 10 TeV, for Galactic cosmic rays injected at the edge of the turbulent superbubble \citep[according to][]{Tolksdorf19}. The different curves correspond to different values of the efficiency parameter $\xi$. The efficient case ($\xi=0.1$) is in blue, the intermediate one ($\xi=1$) in green and the inefficient case ($\xi=10$) in orange. $R_{SB}$ is the superbubble radius and $R_{Esc}$ the free escape boundary radius.}
		\label{fig:CRpatial}
	\end{center}
\end{figure}

Our \g-ray emissivity measurements in the shell are close to the spectral index of 2.6 predicted for an inefficient system with $\xi \gg 1$ (see Fig. \ref{fig:Tobias}). Yet, the superbubble characteristics  \citep[$R_{SB}{\approx} 100$~pc, $B_{int}\approx 1{-}3$~\muG, $v_A{\approx} 25{-}80$~\kmpers, ][\citetalias{Joubaud19}]{Ntormousi17} yield an efficiency $0.2 < \xi(10\,{\rm GeV}/c) < 1.2$ suggesting that CR re-acceleration is possible up to energies of 10-100 GeV if the $L^*$ injection scale of magnetic turbulence is close to 10~pc and the turbulence index approaches $b_T \sim 3/2$. Figure \ref{fig:Tobias} shows that a system with an intermediate efficiency $\xi \sim 1$ would yield a hard spectrum with a spectral index of about 0.7 that is much harder than what we observe. A steeper, Kolmogorov-type of turbulence would be much less efficient in the same conditions ($3 < \xi(10\,{\rm GeV}/c) < 12$). A stronger magnetic field in the hot plasma would lower the $\xi$ ratio. Assuming equipartition with the hot gas pressure \citepalias{Joubaud19}, the field strength of 14~\muG would clearly favour re-acceleration for both types of turbulence scaling ($0.02 \leq \xi(10\,{\rm GeV}/c) \leq 0.3$ for $3/2 \leq b_T \leq 5/3$). 
Substantially reducing the superbubble radius to increase $\xi$ is not supported by the data and the $L^*$ injection scale has only a limited impact on the diffusion coefficient ($D \propto L^{*1/2}$). Therefore, in order to reach an efficiency $\xi(10\,{\rm GeV}/c) {\sim} 10$ that yields a CR spectrum close to our observations, the required turbulence level inside the superbubble should be $\eta_T \lesssim 30\%$ for Kraichnan turbulence. Hence the apparent re-acceleration inefficiency of the superbubble points to a low $\eta_T$ amplitude of the large-scale turbulence and/or to a steep index close to 5/3 of the power density of the fluctuations. Figure \ref{fig:CRpatial} shows that, for inefficient systems like the present one, Galactic CR can penetrate deep into the rarefied medium of the superbubble.

\subsection{Eridu cirrus}

The Eridu cirrus, located at an altitude of about 200-250~pc below the Galactic plane, lies next to, but outside the superbubble. It presents a \g-ray emissivity 34\% lower than the local ISM average, and 28\% lower than the superbubble average. 
We have investigated several leads that could explain a lower CR flux in this cloud. 

First, the loss could result from partial penetration into the cloud. The Eridu cirrus has, however, a modest column density ($1-5 \times10^{20}$~cm$^{-2}$) compared to the nearby superbubble clouds that are pervaded by a larger cosmic-ray flux. \citet{Everett11} studied the evolution of the CR pressure in a diffuse cloud with gas and magnetic-field conditions similar to those in the Eridu cirrus. They used an inner gas density of 100~\percc and a uniform magnetic field of 3~\muG compared to the density of $\sim 7$~\percc and \bsky values of 5~\muG in Eridu \citepalias{Joubaud19}. They found that the Alfven-wave pressure increases by nearly two orders of magnitude at the cloud periphery because of the growth of the streaming instability. But the density of low-energy particles (1 GeV/nucleon) drops by about 5\% only in the cloud interior. There would be even less contrast for higher-energy particles. A 30\% drop as observed in Eridu would require a much larger magnetic-field strength than indicated by the Planck polarisation data in the Eridu cloud and than measured in other diffuse clouds. The CR probed in our analysis should therefore not be depleted because of a lower diffusivity on the cloud boundary.

Clouds of similar structure have already been probed by \citet{PlanckXXVIII15}, with the same analysis method. In their study of the Chamaeleon complex, they identified an intermediate-velocity arc of gas which compares with the Eridu cirrus in several ways: they both have an elongated structure and comparable column densities around $5 \times10^{20}$~cm$^{-2}$; they exhibit large velocity gradients spanning between $-40$ and $-4$~\kmpers in the arc and $-15$ to $-2$~\kmpers in Eridu. The arc lies between 50 and 100~pc below the Galactic plane, assuming the same distances as in  \citet{PlanckXXVIII15}. The arc has, however, a \g-ray emissivity that is 8\% above \qlis for a spin temperature of 140~K, thereby indicating that the lower CR flux in the Eridu cirrus is not linked to the gas filamentary structure and velocity shear.

In order to assess the propagation regime in this cloud, we used the model of \citet{Evoli18} describing the vertical CR transport out of the Galactic disc through the competition of two processes: propagation is dominated by scattering on self-generated Alfvén waves within a few kiloparsecs from the Galactic plane and at sub-TeV energies, whereas ISM turbulence injected by sources near the plane and advected vertically away takes over when it has had time (hence altitude) to cascade down to low enough scales. This model neglects CR diffusion on the toroidal component of the Galactic magnetic field. The Eridu cirrus lies at a distance of 200-250~pc from the Galactic plane and we probed its CR with energies in the range of 2-500 GeV. According to \citet{Evoli18}, we expect them to preferentially diffuse on self-generated Alfvén waves and to be advected at the Alfvén speed.
The estimates of the gas density and magnetic-field strength obtained in \citetalias{Joubaud19} allow to compare the Alfvén speed in the Eridu cirrus ($\gtrsim 4$~\kmpers) and in the \westrim of the superbubble (9-14 \kmpers). We only have an estimate of the \bsky component of the magnetic field in Eridu (hence the lower speed limit given above), but having an Alfvén velocity in the cirrus much larger than in the \westrim would require an unusually large total field strength in excess of $>10-20$~\muG for this type of diffuse cloud \citep{Crutcher12}. So the slower transport by Alfvén waves in the Eridu cirrus cannot explain a loss of CR. 

On the other hand, dust polarisation observations with \Planck show both a low dispersion in polarisation orientations and a high polarisation fraction along the cirrus, as along the outer rim of the superbubble. This is visible 
in Fig.2 of \citet{Soler18}. The polarisation data implies a low level of magnetic turbulence inside the cirrus, of order $\delta B/B_0 \sim 0.1$, which compares with the low level found along the superbubble rim due to magnetic compression in the expanding shell. 
A closer view of the magnetic orientation towards the Eridu cirrus is displayed in Fig. \ref{fig:bfield}. Its magnetic field appears to be ordered and oriented along the filamentary structure of the cirrus, as expected for low-column density atomic clouds \citep{Clark14, PlanckXXXII2016}. An ordered magnetic field with a low level of turbulence favours a rapid transport of CR along the field lines.

Figure \ref{fig:bfield} further illustrates the change in polarisation orientation between the cirrus direction and the superbubble \westrim on the one hand, and between the cirrus and the rest of the gas at very low negative latitudes where the field is mostly parallel to the Galactic plane. Both changes in Stokes parameters are significant at the 40\arcmin scale and they happen in regions of high signal-to-noise ratio in polarised intensity ($P/\sigma_P \geq 3$). The dust signal from the cirrus dominates the  polarisation data along those lines of sight. This data therefore suggests that the Eridu cirrus is on a different bundle of field lines. It does not follow the circular rim of the superbubble shell, nor the plane-parallel configuration of the diffuse \hi layer, but the field lines rather point out towards lower Galactic latitudes, with a $\sim 45$\degr\ angle from the south Galactic pole. The lower CR flux pervading the Eridu cirrus could then be due to propagation along a magnetic `tube' pointing towards the Galactic halo. We are currently studying another cirrus with a similar orientation towards the halo to confirm this possibility. 

We also note that the older part of the superbubble corresponds to a direction in the sky of low synchrotron intensity at 408 MHz \citep{Haslam82}. This emission probes CR electrons with energies near 10~GeV in the magnetic fields of the cirrus and superbubble shells. The LIS electron spectrum has a spectral index of $-3.5$ around this energy \citep{Boschini18}. Despite the enhanced field strengths in the superbubble shell, we observe the same synchrotron intensities in the Eridu and \westrim directions for which we have magnetic field measurements \citepalias{Joubaud19}. The ratios of synchrotron intensities between the Eridu and \westrim directions indicate that the CR electron flux in the cirrus is half that in the superbubble \westrim, thereby confirming that there are fewer CR in this diffuse cloud than along the superbubble shells.  

\begin{figure}[ht]
	\begin{center}
		\resizebox{\hsize}{!}{\includegraphics{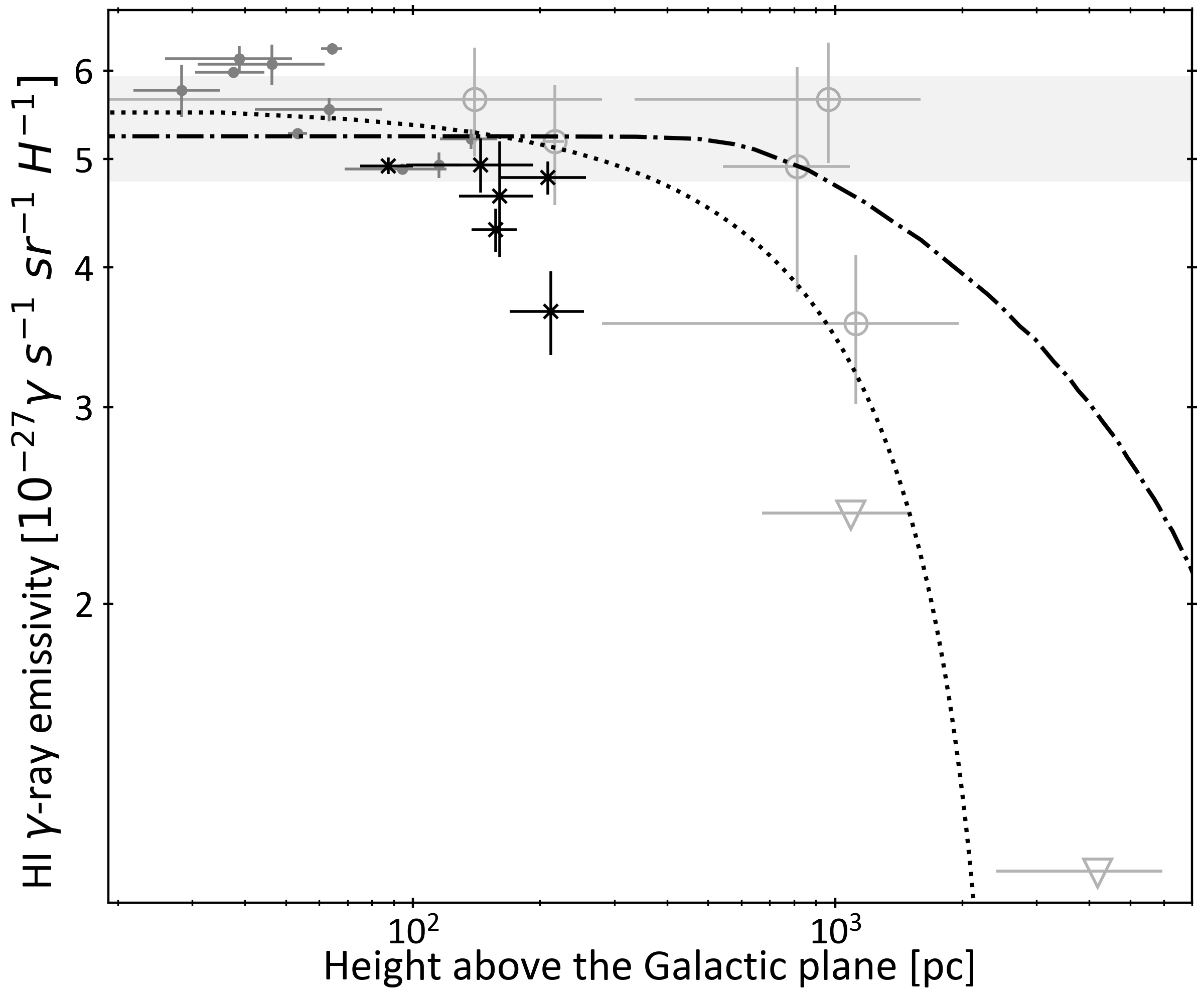}}
		\caption{Integrated (0.4-10 GeV) \g-ray emmissivity per gas nucleon as a function of height above the Galactic plane. The black data points are from this study, the dark grey points from nearby anticentre clouds \citep{Remy17} and the light grey ones from high-latitude clouds \citep[adapted from][ see text]{Tibaldo15}. The grey band gives the $\pm1\sigma$ dispersion around the mean emissivity found at low Galactic height ($\leq 300$~pc). The dotted curve shows the best fit for pure CR diffusion and advection with Galactic height (see text). The dash-dotted curve corresponds to the distribution of 10~GeV protons in the halo model of \citet{Evoli18}, rescaled to the \qlis emissivity at low Galactic height.}
		\label{fig:Zgal} 
	\end{center}
\end{figure}
    
\subsection{Local cosmic-ray vertical gradient}

Assuming there is no strong confinement in the superbubble, one could consider the simplest 1D model of pure diffusion perpendicular to a thin Galactic disc, with a given $\pm H$ halo height where the CR flux vanishes \citep[see for example the review by][]{Amato18}. The CR steady-state distribution function linearly drops with altitude up to the halo height: $f(z,p)= f_0(p)(1 - |z|/H)$. For a uniform diffusion coefficient $D(p)$ and a $Q_0(p)$ spectrum injected by the sources, the CR spectrum is uniform and softened by diffusion as $f_0(p) \propto Q_0(p)/D(p)$. 
A slightly more refined description adds a $\beta$ index to the spatial dependence,  $f \propto 1- (|z|/z_{\rm{max}})^{\beta}$, as done by \citet{Ackermann12} with the set of GALPROP models, yielding $\beta$ values ranging from 1 to 1.5. A third simple case assumes no losses, uniform diffusion perpendicular to the Galactic disc and vertical advection at a velocity $v_{a+w}$ that combines the Alfvén speed (e.g. for CR diffusion on self-generated Alfvén waves) and a wind velocity (for hot-gas pressure driven outflow of the background plasma). The CR distribution function then scales as $f(z,p) = f_0(p) [1 - e^{-\zeta (1-|z|/H)}]/[1 - e^{-\zeta}]$, with $\zeta = v_{a+w} H / D(p)$.

In order to probe the relationship between altitude and CR density, we plotted in Fig. \ref{fig:Zgal} the \g-ray emissivities measured in several clouds against the altitude to the Galactic plane. The sample includes the local values found by \citet{Remy17} in anticentre clouds. We have discarded their Perseus cloud estimate as they note that strong confusion between \hi and CO structures in this compact cloud leads to an unreliable emissivity measurement in each phase. The sample also includes measurements towards high-latitude clouds seen at various distances, including intermediate- and high-velocity \hi clouds \citep{Tibaldo15}. The distance to the nearby Ursa Major cloud included in this analysis has recently been constrained to be 360 $\pm$ 20 pc, so we have updated its height in Fig. \ref{fig:Zgal}. The distances to the high-velocity clouds are still poorly constrained. Figure \ref{fig:Zgal} shows that the emissivities found in the superbubble shells are consistent with the spread shown in local clouds from previous studies. Systematic uncertainties in \hi optical depth corrections can explain the 10-15\% dispersion, but the Eridu cirrus stands at variance as the lowest black data point. 

The overall trend in Fig. \ref{fig:Zgal} suggests an emissivity drop with Galatic height. A linear fit to the sample of local clouds ($|z|\leq 300$~pc) yields a 5.5$\sigma$ detection of a slope. The resulting halo size of $490\pm60$~pc (including the Eridu data point or not) is too low compared to the $5.4\pm 1.4$~kpc value inferred from GALPROP models \citep{Trotta11}. It compares favourably with the few hundred parsec heights favoured by dynamic spiral arm models, which add ISM advection downstream of the spiral arms to diffusion \citep{Nava17}. A linear fit to the set of high-latitude data points (only) yields a halo height of 2.2$\pm$1.9~kpc \citep{Tibaldo15}. Figure \ref{fig:Zgal} also shows the best-fit $f(z)$ curve obtained for the case combining vertical diffusion and advection (dotted curve). It yields a halo height of $2.6\pm 0.1$~kpc in reasonable agreement with elemental constraints, and a small scaling factor $\zeta=(1.7 \pm 0.3)\times 10^{-3}$. The latter implies that, for a standard diffusion coefficient of $6.2 \times 10^{28}$~cm$^2$~s$^{-1}$~$(R/10\,{\rm GV})^{0.344}$ \citep{Gully19}, vertical advection has little impact on the local propagation of 10~GeV CR. Adding 10\% systematic uncertainties on the data points because of \hi optical depth corrections significantly impacts the halo height which becomes $4.5\pm 0.2$~kpc, but it does not change the conclusion about a negligible advection velocity. As discussed in the previous section, the loss of CR in the Eridu cirrus may correspond to its specific magnetic configuration, and not to large-scale transport properties. Removing this cloud from the sample does not significantly change the halo height indicated by the fit. Additional \g-ray analyses of nearby clouds at an altitude of about 500~pc are clearly necessary to confirm this trend.

Figure \ref{fig:Zgal} also displays the expected CR profile from the halo model of \citet{Evoli18} for 10-GeV protons. We have set their distribution to match the \g-ray emissivity at low altitude. Their model includes advection of ISM turbulence away from the plane, as well as CR scattering on self-generated waves induced by the streaming instability through the vertical CR gradient. This model successfully reproduces the hardening of primary CR near 300 GV, but Figure \ref{fig:Zgal} shows that the modelled drop in CR flux with height is shallower than suggested by the \g-ray observations of individual clouds.

\section{Conclusion}

In order to assess the impact of the turbulent environment of superbubbles on CR propagation, we probed \g-ray emissivities in atomic clouds of the Orion-Eridanus superbubble using ten years of \Fermi-LAT data. Our analysis revealed a CR spectrum that is uniform within 6\% between the superbubble shell and the average found in the rest of the local ISM within a few hundreds parsecs from the Sun. The CR spectrum in the superbubble is also consistent with the measurement in the Solar system. Uncertainties from the HI optical depth correction in those shells amount to 7\% and can easily explain such a small difference. We found no spectral evidence of CR (re)-acceleration even though the stellar and gas content of the superbubble suggests that it could actively re-accelerate CR. Its inefficiency could be due to a lower frequency and space density of supernovae in the last Myr, compared to other starburst regions like Cygnus X. It could also be due to a weak average magnetic field ($\lesssim 1$~\muG) inside the superbubble or to a steep, Kolmogorov-like spectrum of the magnetic turbulence if the average field reaches a few microGauss.

We have gathered a sample of nearby clouds that have been probed in \g rays with Fermi-LAT using the same analysis method to separate the clouds along the lines of sight and to separate the different gas phases in order to measure the \g-ray emissivity in the atomic phase. Their comparison shows emissivity variations as a function of height above the Galactic plane. Adding more clouds to this sample will allow for a local estimation of the CR halo size and will establish or rule out the importance of vertical advection, rather than inferring these important parameters from larger-scale propagation models.

A cirrus cloud located outside the superbubble presents a 34\% lower CR flux compared to the local ISM average. Its filamentary structure lies along magnetic-field lines that appear to be highly inclined with respect to the Galactic plane. This result opens the possibility to investigate how CR flow along magnetic flux tubes in order to better understand the role of gas fountains and of their magnetic field loops in CR escape to the halo.

\begin{acknowledgements}
We acknowledge the financial support from the ANR-DFG grant CRiBs (ANR-15-CE31-0019-01 and Schl 201/34-1) for this work. We thank Alexei Ivlev and Andrei Bykov for stimulating discussions and the referee for his/her suggestions about the turbulence. The \Fermi LAT Collaboration acknowledges generous ongoing support from a number of agencies and institutes that have supported both the development and the operation of the LAT as well as scientific data analysis. These include the National Aeronautics and Space Administration and the Department of Energy in the United States, the Commissariat \`a l'Energie Atomique and the Centre National de la Recherche Scientifique / Institut National de Physique Nucl\'eaire et de Physique des Particules in France, the Agenzia Spaziale Italiana and the Istituto Nazionale di Fisica Nucleare in Italy, the Ministry of Education, Culture, Sports, Science and Technology (MEXT), High Energy Accelerator Research Organisation (KEK) and Japan Aerospace Exploration Agency (JAXA) in Japan, and the K.~A.~Wallenberg Foundation, the Swedish Research Council and the Swedish National Space Board in Sweden. Additional support for science analysis during the operations phase is gratefully acknowledged from the Istituto Nazionale di Astrofisica in Italy and the Centre National d'\'Etudes Spatiales in France. This work performed in part under DOE Contract DE-AC02-76SF00515.

\end{acknowledgements}

\bibliographystyle{aa}
\bibliography{References}

\appendix

\section{\hi spin temperature}
\label{ap:Tspin}

The evolution of the maximum log-likelihood of the \g-ray fit can be used to constrain the choice of average \hi spin temperature that best matches the \nhi column density maps of the different clouds to the structure of the \g rays produced by CR interactions in the atomic gas. The DNM maps were iteratively derived for each spin temperature value independently. We find that the fit quality is best for a spin temperature of 100~K. We did not attempt to use different spin temperatures for the different clouds present in the analysis region.

 \begin{figure}[h]
	\begin{center}
		\resizebox{\hsize}{!}{\includegraphics{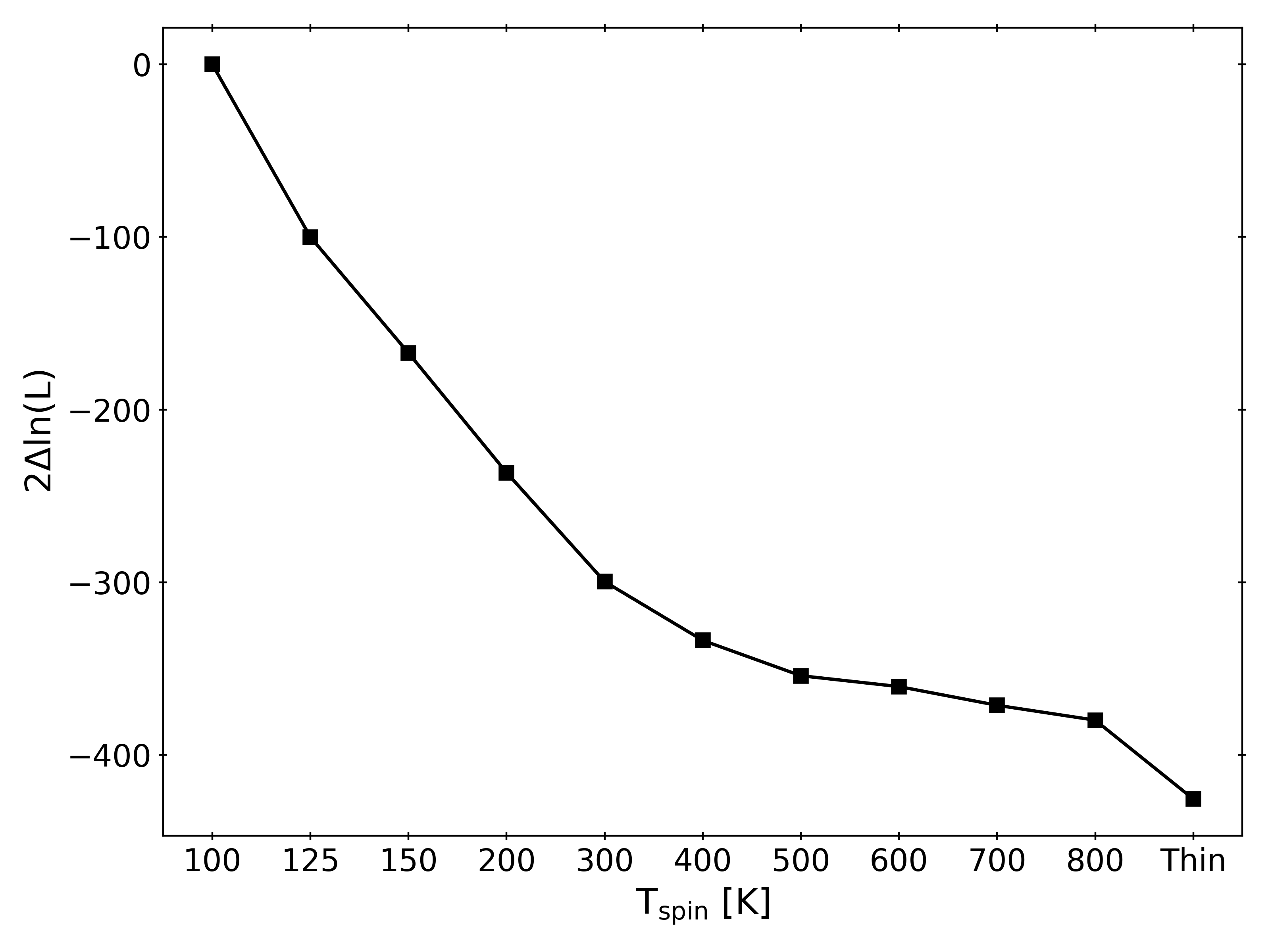}}
		\caption{Evolution of the log-likelihood ratio of the \g-ray fit as a function of the spin temperature used to calculate the $N_{\mathrm{\hi}}$ column densities.}
		\label{fig:LnL} 
	\end{center}
\end{figure}

\section{Jackknife tests}
\label{ap:Jack}

Repeating the last dust and \g-ray fits a thousand times over subsets of the analysis region, after masking out 20\% of the pixels, allowed us to evaluate the systematic uncertainties in our linear modelling approximations. In \g rays, the jackknife tests have been performed only for the total 0.25–63 GeV energy band.

\begin{figure*}[ht]
	\centering
	\includegraphics[width=\hsize]{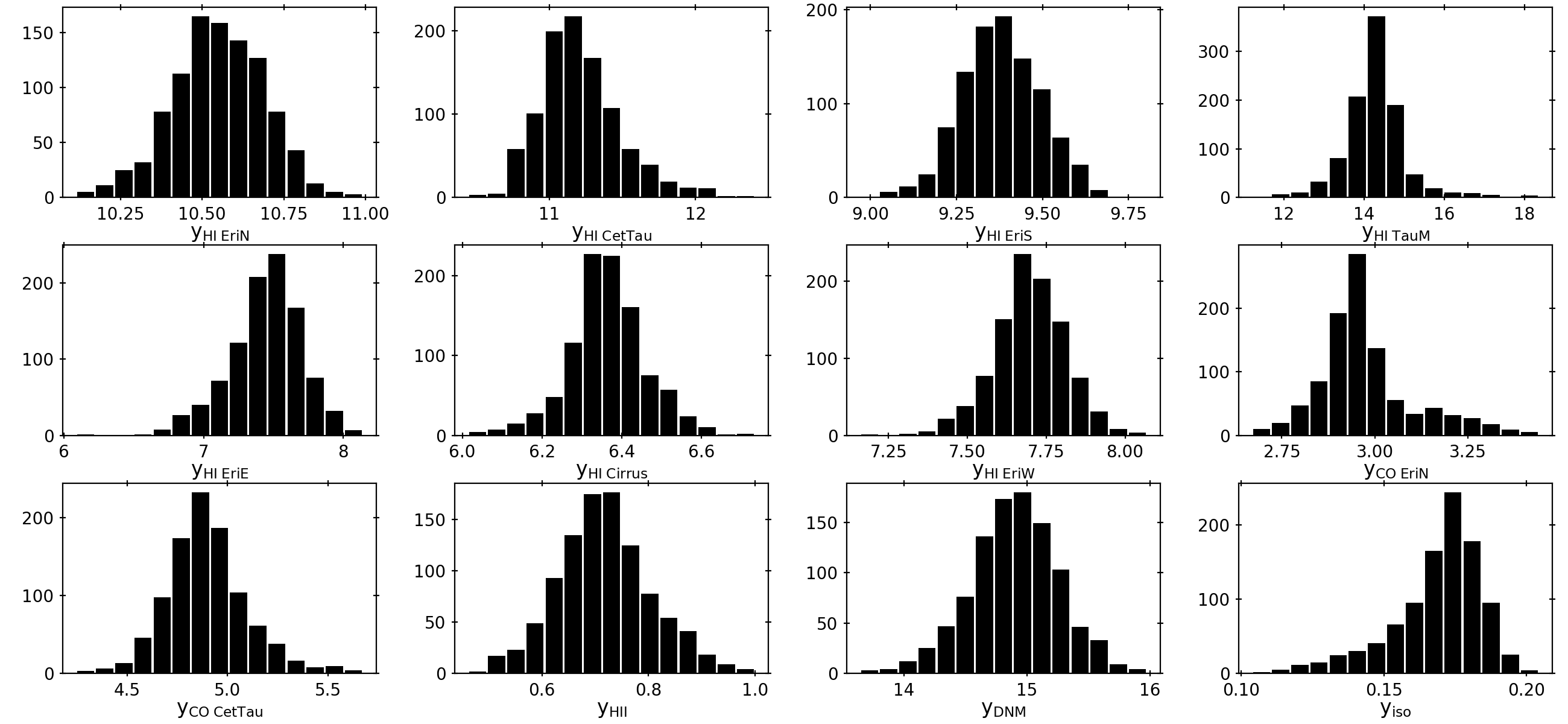}
	\caption{Number distribution of the dust model coefficients over 1000 jackknife fits for an \hi spin temperature of 100 K. \yhi and \ydnm are in units of \yhTunit, \yco in \ycoTunit, \yhii and \yiso in \yisoTunit.}
	\label{fig:JackDust}
\end{figure*}

\begin{figure*}[ht]
	\centering
	\includegraphics[width=\hsize]{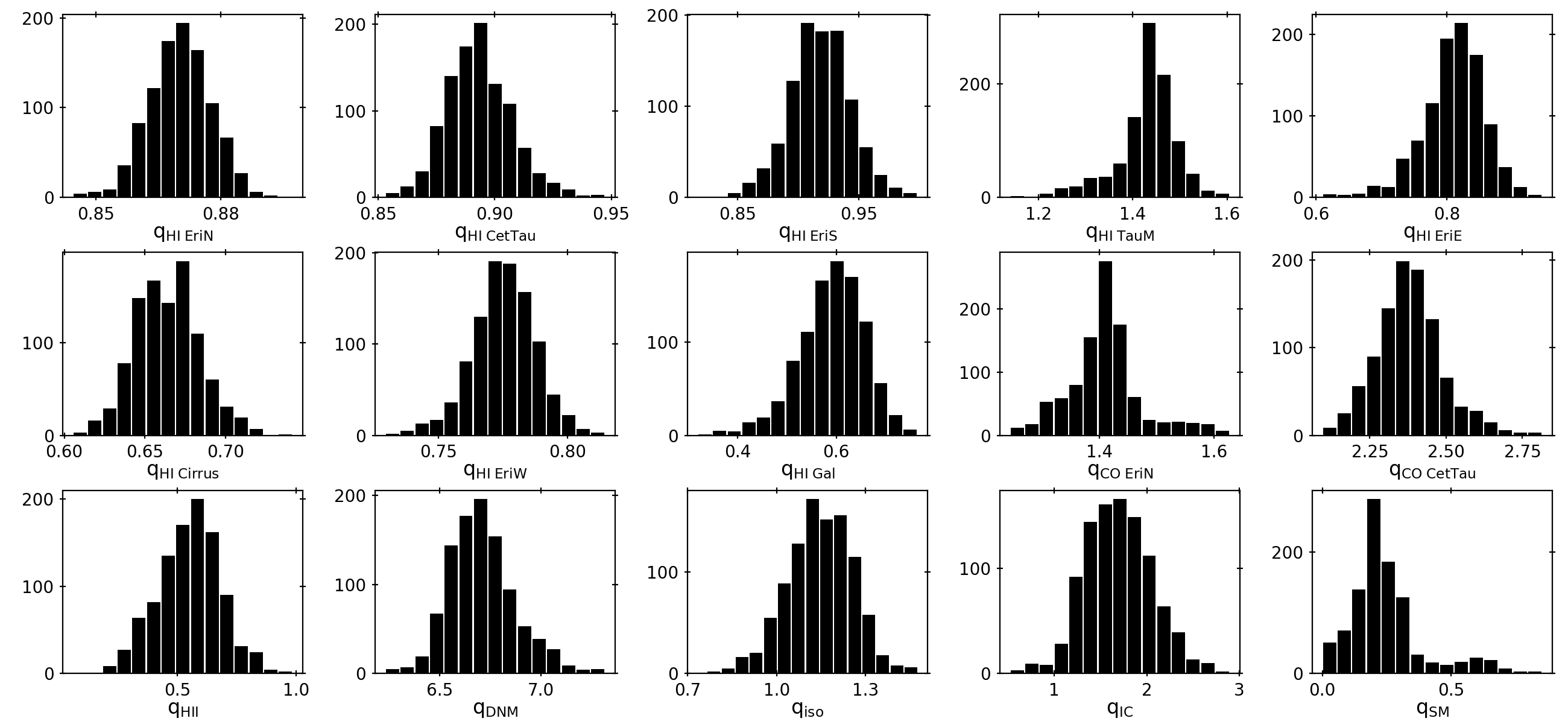}
	\caption{Number distribution of the \g-ray model coefficients obtained for 1000 jackknife fits for an \hi spin temperature of 100 K. \qco are in units of \qcounit, \qhii in \qhiiunit, \qdnm in \qdnmTunit. The other q coefficients are simple normalisation factors.}
	\label{fig:JackGam}
\end{figure*}

\section{Best fit coefficients of the \g-ray and dust models}
\label{ap:coef}

The best fit corresponds to an \hi spin temperature of 100~K. Errors combine statistical uncertainties and the results of the jackknife tests.

\setlength{\tabcolsep}{5pt}
\begin{sidewaystable} 
\centering 
\begin{tabular}{ l | c c c c c c c c || l | c } 
 \hline \hline	
$\gamma$-ray & \multicolumn{8}{|c||}{Energy band [MeV]} & \multicolumn{2}{c|}{$ \tau_{353}$} \\ 
 model & $10^{ 2.4 }$-$10^{ 2.6 }$ & $10^{ 2.6 }$-$10^{ 2.8 }$ & $10^{ 2.8 }$-$10^{ 3.0 }$ & $10^{ 3.0 }$-$10^{ 3.2 }$ & $10^{ 3.2 }$-$10^{ 3.6 }$ & $10^{ 3.6 }$-$10^{ 4.0 }$ & $10^{ 4.0 }$-$10^{ 4.8 }$ & $10^{ 2.4 }$-$10^{ 4.8 }$ & \multicolumn{2}{c|}{model} \\ 
 \hline 
q$_{\mathrm{ HI\ EriN } }$ &      0.863 $\pm$      0.015 &      0.873 $\pm$      0.013 &      0.870 $\pm$      0.013 &      0.871 $\pm$      0.014 &      0.856 $\pm$      0.015 &      0.879 $\pm$      0.030 &      0.857 $\pm$      0.070 &     0.8689 $\pm$     0.0090 & y$_{\mathrm{ HI\ EriN } }$ &      1.056 $\pm$      0.014 \\[0.1cm] 
q$_{\mathrm{ HI\ EriS } }$ &      0.951 $\pm$      0.062 &      0.886 $\pm$      0.050 &      0.901 $\pm$      0.048 &      0.893 $\pm$      0.051 &      0.952 $\pm$      0.052 &       1.04 $\pm$       0.11 &       0.86 $\pm$       0.26 &      0.918 $\pm$      0.032 & y$_{\mathrm{ HI\ EriS } }$ &      0.939 $\pm$      0.013 \\[0.1cm] 
q$_{\mathrm{ HI\ EriE } }$ &       0.73 $\pm$       0.11 &       0.91 $\pm$       0.10 &      0.869 $\pm$      0.097 &       0.94 $\pm$       0.11 &       0.75 $\pm$       0.10 &       0.58 $\pm$       0.23 &       0.65 $\pm$       0.58 &      0.826 $\pm$      0.063 & y$_{\mathrm{ HI\ EriE } }$ &      0.749 $\pm$      0.028 \\[0.1cm] 
q$_{\mathrm{ HI\ EriW } }$ &      0.748 $\pm$      0.046 &      0.848 $\pm$      0.035 &      0.742 $\pm$      0.032 &      0.750 $\pm$      0.035 &      0.808 $\pm$      0.036 &      0.691 $\pm$      0.079 &       0.70 $\pm$       0.19 &      0.777 $\pm$      0.019 & y$_{\mathrm{ HI\ EriW } }$ &      0.776 $\pm$      0.012 \\[0.1cm] 
q$_{\mathrm{ HI\ Cirrus } }$ &      0.651 $\pm$      0.073 &      0.616 $\pm$      0.056 &      0.693 $\pm$      0.054 &      0.700 $\pm$      0.059 &      0.668 $\pm$      0.061 &       0.56 $\pm$       0.14 &       0.00 $\pm$       0.32 &      0.657 $\pm$      0.031 & y$_{\mathrm{ HI\ Cirrus } }$ &      0.638 $\pm$      0.011 \\[0.1cm] 
q$_{\mathrm{ HI\ CetTau } }$ &      0.946 $\pm$      0.038 &      0.877 $\pm$      0.030 &      0.908 $\pm$      0.028 &      0.858 $\pm$      0.029 &      0.872 $\pm$      0.029 &      0.841 $\pm$      0.059 &       0.82 $\pm$       0.13 &      0.892 $\pm$      0.018 & y$_{\mathrm{ HI\ CetTau } }$ &      1.129 $\pm$      0.029 \\[0.1cm] 
q$_{\mathrm{ HI\ Gal } }$ &       0.55 $\pm$       0.16 &       0.45 $\pm$       0.13 &       0.58 $\pm$       0.13 &       0.62 $\pm$       0.14 &       0.54 $\pm$       0.14 &       0.66 $\pm$       0.30 &       0.41 $\pm$       0.63 &      0.606 $\pm$      0.086 &  & \\[0.1cm] 
q$_{\mathrm{ CO\ EriN } }$ &       1.58 $\pm$       0.10 &      1.520 $\pm$      0.089 &      1.359 $\pm$      0.081 &      1.347 $\pm$      0.084 &      1.259 $\pm$      0.081 &       1.35 $\pm$       0.14 &       0.82 $\pm$       0.26 &      1.426 $\pm$      0.070 & y$_{\mathrm{ CO\ EriN } }$ &       2.99 $\pm$       0.14 \\[0.1cm] 
q$_{\mathrm{ CO\ CetTau } }$ &       2.34 $\pm$       0.54 &       2.40 $\pm$       0.42 &       2.44 $\pm$       0.37 &       2.43 $\pm$       0.36 &       2.15 $\pm$       0.32 &       2.00 $\pm$       0.60 &        2.6 $\pm$        1.4 &       2.41 $\pm$       0.21 & y$_{\mathrm{ CO\ CetTau } }$ &       4.90 $\pm$       0.27 \\[0.1cm] 
q$_{\mathrm{ HII } }$ &       0.51 $\pm$       0.31 &       0.81 $\pm$       0.28 &       0.70 $\pm$       0.25 &       0.51 $\pm$       0.23 &       0.00 $\pm$       0.18 &       0.00 $\pm$       0.42 &        0.0 $\pm$        1.0 &       0.56 $\pm$       0.16 & y$_{\mathrm{ HII } }$ &      0.708 $\pm$      0.095 \\[0.1cm] 
q$_{\mathrm{ DNM } }$ &       6.81 $\pm$       0.29 &       6.67 $\pm$       0.26 &       6.73 $\pm$       0.26 &       6.69 $\pm$       0.27 &       6.69 $\pm$       0.28 &       7.05 $\pm$       0.53 &        6.7 $\pm$        1.1 &       6.77 $\pm$       0.19 & y$_{\mathrm{ DNM } }$ &      1.490 $\pm$      0.040 \\[0.1cm] 
q$_{\mathrm{ Iso } }$ &       1.16 $\pm$       0.23 &       1.05 $\pm$       0.25 &       1.40 $\pm$       0.32 &       1.41 $\pm$       0.41 &       1.17 $\pm$       0.41 &       1.14 $\pm$       0.40 &       0.73 $\pm$       0.36 &       1.17 $\pm$       0.15 & y$_{\mathrm{ Iso } }$ &      0.163 $\pm$      0.017 \\[0.1cm] 
q$_{\mathrm{ IC } }$ &       1.12 $\pm$       0.75 &       2.03 $\pm$       0.83 &       1.60 $\pm$       0.84 &       1.82 $\pm$       0.96 &       2.22 $\pm$       0.93 &        1.6 $\pm$        1.3 &        2.6 $\pm$        1.9 &       1.64 $\pm$       0.47 & \\[0.1cm] 
 q$_{\mathrm{ SM } }$ &       0.33 $\pm$       0.48 &       0.41 $\pm$       0.43 &       0.00 $\pm$       0.31 &       0.72 $\pm$       0.51 &       0.58 $\pm$       0.40 &       0.00 $\pm$       0.27 &       0.42 $\pm$       0.50 &       0.25 $\pm$       0.20 & \\[0.1cm] 
 \hline \hline 
\end{tabular} 
\caption{Best-fit coefficients for the \g-ray and dust fits.The \qco coefficients are expressed in \qcounit, \qdnm in \qdnmTunit. $q_{\rm{iso}}$, $q_{\rm{IC}}$ and $q_{\rm{SM}}$ are normalisation factors. yhi and \ydnm are expressed in \yhTunit, \yco in \ycoTunit, \yhii and \yiso in \yisoTunit.}
\end{sidewaystable}

\section{Normalised \g-ray emissivities}
\label{ap:em}

The \g-ray emissivity spectra normalised by the local interstellar medium average, \qlis, make it possible to highlight any spectral or flux variation. They were derived for a spin temperature of 100~K, favoured by the fit, and for the  optically thin case giving an upper limit to the emissivity. The small variations between the two cases show the limited influence of the \hi optical depth correction on our results.

 \begin{figure*}[h]
	\begin{center}
		\resizebox{\hsize}{!}{\includegraphics{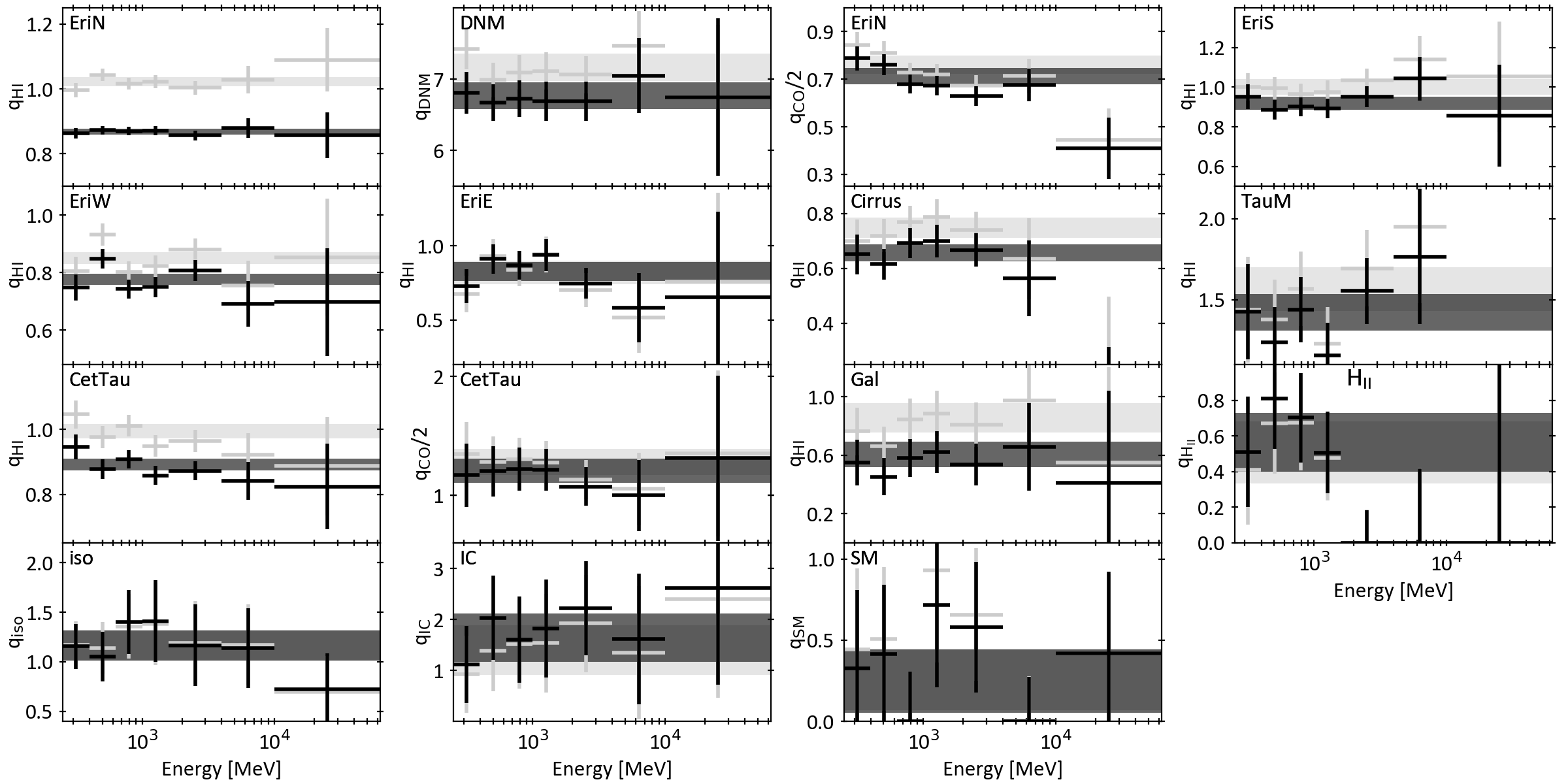}}
		\caption{Spectral evolution of the \g-ray emissivities relative to the local interstellar medium \qlis. The light grey crosses correspond to the optically thin spectrum and the black crosses to the 100~K case, favoured by the fit. The corresponding bands give the average emissivity and its errors. The \qco coefficients are expressed in \qcounit, \qhii in \qhiiunit, \qdnm in \qdnmTunit. The other $q$ coefficients are simple normalisation factors.}
		\label{fig:qqLISmos} 
	\end{center}
\end{figure*}

\end{document}